\documentclass[a4paper, onecolumn, 11pt, unpublished]{quantumarticle}
\pdfoutput=1
\usepackage[numbers, sort&compress]{natbib}
\usepackage[utf8]{inputenc}
\usepackage[english]{babel}
\usepackage[T1]{fontenc}
\usepackage{amsmath,amssymb,amsthm}
\usepackage{hyperref}
\usepackage{dsfont}
\usepackage{mathrsfs}
\usepackage{enumitem}

\usepackage{graphicx}
\usepackage{color}
\usepackage{upgreek}
\usepackage{float}
\usepackage{lipsum}
\usepackage{mathrsfs}
\usepackage{booktabs}
\usepackage{algorithm} 
\usepackage{algpseudocode}
\usepackage{acronym}

\newcommand{\ket}[1]{\ensuremath{\left| #1 \right\rangle}}
\newcommand{\bra}[1]{\ensuremath{\left\langle #1 \right|}}

\newcommand{\Tr}[0]{\ensuremath{\text{Tr}}}

\AtBeginDocument{
\heavyrulewidth=.08em
\lightrulewidth=.05em
\cmidrulewidth=.03em
\belowrulesep=.65ex
\belowbottomsep=0pt
\aboverulesep=.4ex
\abovetopsep=0pt
\cmidrulesep=\doublerulesep
\cmidrulekern=.5em
\defaultaddspace=.5em
}
\begin{document}

\title{
Practical learning of multi-time statistics in open quantum systems 
}

\author{G. A. L. White}
\email{greg.al.white@gmail.com}
\affiliation{Dahlem Center for Complex Quantum Systems, Freie Universit\"at Berlin, 14195 Berlin, Germany}
\affiliation{School of Physics and Astronomy, Monash University, Clayton, VIC 3800, Australia}
\affiliation{School of Physics, University of Melbourne, Parkville, VIC 3010, Australia}


\author{L. C. L. Hollenberg}
\affiliation{School of Physics, University of Melbourne, Parkville, VIC 3010, Australia}

\author{C. D. Hill}
\email{charles.hill1@unsw.edu.au}
\affiliation{Silicon Quantum Computing, The University of New South Wales, Sydney, New South Wales 2052, Australia}
\affiliation{School of Physics, University of Melbourne, Parkville, VIC 3010, Australia}
\affiliation{School of Mathematics and Statistics, University of Melbourne, Parkville, VIC, 3010, Australia}

\author{K. Modi}
\email{kavan@quantumlah.org}
\affiliation{Science, Mathematics and Technology Cluster, Singapore University of Technology and Design\\ 8 Somapah Road, 487372 Singapore}
\affiliation{School of Physics and Astronomy, Monash University, Clayton, VIC 3800, Australia}
           
\begin{abstract}
Randomised measurements can efficiently characterise many-body quantum states by learning the expectation values of observables with low Pauli weights. In this paper, we generalise the theoretical tools of classical shadow tomography to the temporal domain to explore multi-time phenomena. This enables us to efficiently learn the features of multi-time processes such as correlated error rates, multi-time non-Markovianity, and temporal entanglement. We test the efficacy of these tools on a noisy quantum processor to characterise its noise features. Implementing these tools requires mid-circuit instruments, typically slow or unavailable in current quantum hardware. We devise a protocol to achieve fast and reliable instruments such that these multi-time distributions can be learned to a high accuracy. This enables a compact matrix product operator representation of large processes allowing us to showcase a reconstructed 20-step process (whose naive dimensionality is that of a 42-qubit state). Our techniques are pertinent to generic quantum stochastic dynamical processes, with a scope ranging across condensed matter physics, quantum biology, and in-depth diagnostics of noisy intermediate-scale quantum devices.
\end{abstract}

\maketitle

\section{Introduction}

There has in recent years been a significant increase in both accessibility to and interest in quantum information processors, including computing devices, sensors, and communication tools. Additionally, related fields such as condensed matter physics and biology are increasingly investigating the direct impacts of quantum mechanics within their areas.
The dynamics of open quantum systems play a crucial role across these fields, acting as sources of complex noise, hidden signals, and even foundational physical insights. Despite the growing potential of this technology to study open quantum system dynamics, its capacity remains largely underdeveloped. Recent developments in the theory of quantum stochastic processes have closed this gap to map many-time processes to many-body states~\cite{chiribella_memory_2008, Pollock2018a, 1367-2630-18-6-063032, PhysRevA.98.012328, Milz2021PRXQ}. It is becoming evident that many-time physics is as vibrant as many-body physics; they are endowed with nontrivial temporal entanglement~\cite{aharonov2009multiple, milz21}, exotic causal properties~\cite{chiribella_quantum_2013, OreshkovETAL2012, Ringbauer2017, Carvacho2017, Milz2018, Ringbauer-npjQI,procopio2015experimental}, rich statistical structures~\cite{Romero2020}, resources~\cite{berk, berk2021extracting, arXiv:2110.03233,araujo2014computational,taddei2019quantum}, and a window into novel physics~\cite{hardy_operator_2012, cotler_superdensity_2017}. Yet, experimental exploration of large-scale many-time physics remains missing. This is because of the multitude of technical problems stemming from experimental limitations, data sizes, and subtle differences between many-body and many-time states.

In this paper, we show that, despite current limitations in hardware control, this challenge can be met through methodological and conceptual advancements. Specifically, we present a new toolkit designed to uncover temporal correlations within general open quantum systems. We leverage classical shadow tomography—adapted here for processes—to analyse noise and simulate processes in quantum devices, characterizing a 20-step process and extracting a range of properties from it. We approach this work with the view that quantum correlations are fundamental to numerous intriguing microscopic phenomena, where systems exhibit connections stronger than any classical framework can describe. They impact the macroscopic properties of many-body quantum systems, leading to remarkable phases of matter. Discussions on this topic often focus on spatial correlations observed at a specific moment, or where time merely functions as a parameter in the exploration of spatial characteristics. In contrast, individual quantum systems can exhibit complex dynamics, especially when interacting with their environment, leading to intricate many-time (quantum) correlations~\cite{milz21, White-MLPT, aloisio-complexity, MTP1}. Addressing these correlations is essential for advancing toward fault-tolerant quantum computers. In other technologies, such as quantum sensors, these correlations have the potential to enhance performance, including sensitivity. 


Specifically, in this work, we address the question: how can many-time phenomena, as described above, be accessed in quantum experiments and on near-term quantum hardware? In doing so, we pave the way for the realistic observation and study of temporal correlations, placing them on a similar level as many-body states. To achieve this, we combine established approaches in many-body physics with new techniques specifically designed for observing temporal quantum states, particularly under various control paradigms. We show that quantum computers, even in their current state, offer a timely opportunity to explore the richness of many-time physics. While these machines are too noisy to manipulate into the large entangled registers required for interesting quantum algorithms, they are riddled with complex noise~\cite{White-NM-2020,White-MLPT,MTP1}, which we can access with high-fidelity controls. Thus, we show that there is scope to explore interesting physics even with relatively small devices -- a single qubit probed across many times, for example, can exhibit many of the same properties as large quantum states. In this sense, these methodological advances show that highly nontrivial many-time physics can be more accessible than many-body physics for the near-term quantum processors, which opens rich new areas of investigation. Applications of this work range from the diagnostics of temporally correlated noise in quantum technologies~\cite{altherr2021quantum, White-NM-2020}, to the observation and simulation of dynamical complexity in condensed matter systems~\cite{nitzan2003electron, PhysRevLett.124.043603, PhysRevLett.115.043601}, to quantum biology~\cite{lambert2013quantum, McGuinness2011} and quantum causal modelling~\cite{1367-2630-18-6-063032}.

We begin with a brief introduction to quantum stochastic processes and many-time correlations in Sec.~\ref{sec:background}. Then we present our main theoretical results of generalising the classical shadow tomography to the spatiotemporal realm in Sec.~\ref{sec:shadow}. To implement classical spatiotemporal shadow tomography requires control that is typically available in current quantum hardware, namely mid-circuit measurements. We overcome this limitation in Sec.~\ref{sec:IC-control}. We then demonstrate, in Sec.~\ref{sec:correlated-noise}, the high efficacy of these methods by extracting features of many-time physics in a series of quantum simulations of complex dynamics on IBM Quantum processors. In particular, we characterise a 20-step quantum process and extract temporal correlations from it. Our conclusions are presented in Sec.~\ref{sec:conc}

\section{Background on Quantum Stochastic Processes}
\label{sec:background}

In any experiment, there exists a distinction between the controllable operations executed by the experimenter on a $d$-dimensional system and the uncontrollable stochastic dynamics arising from system-environment ($SE$) interactions. A quantum stochastic process captures all quantum correlations across multiple times, denoted $\mathbf{T}_k := {t_0, \cdots, t_k}$. On a quantum computer, this could represent a sequence of gates; in a sensor, it might correspond to a preparation protocol. Understanding the structure of non-Markovian quantum noise has been a significant challenge, presenting a serious practical obstacle to developing quantum information processors. However, recent advancements have mitigated some of these theoretical challenges, enabling the practical characterization of complex noise, including many-time correlations.


We begin with a brief review of the theory of quantum stochastic processes. Key concepts include completely positive operations, process tensors, Choi representations, the Stinespring dilation, instruments, testers, and more. We start by revisiting quantum channels, which represent two-time correlations. After reintroducing these foundational concepts, we will extend them to the many-time setting.

\subsection{Quantum channels}
\label{app:channels}

\begin{figure}
    \centering
    \includegraphics[width=\linewidth]{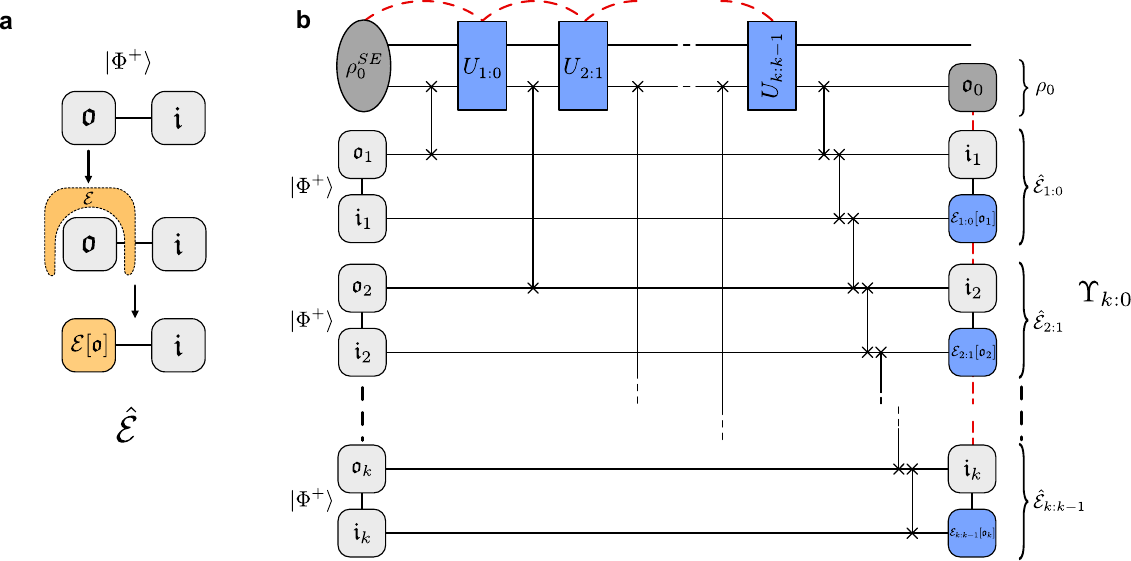}
    \caption{Circuit diagrams of the conventional and generalised Choi-Jamiolkowski isomorphism. \textbf{a} Two-time processes are represented by quantum channels; their Choi state is given by the channel acting on one half of a Bell pair. \textbf{b} many-time processes are represented by process tensors; their Choi state is given for a $k$-step process by swapping in one half of $k$ Bell pairs at different times. The non-Markovian correlations are mapped onto spatial correlations in the Choi state. The marginals of the process tensor are quantum channels, as well as the average initial state.}
    \label{fig:choi_reps}
\end{figure}

A completely positive (CP) channel is the mapping from bounded linear operators, $\mathscr{B}(\mathcal{H}_{\text{in}})\rightarrow \mathscr{B}(\mathcal{H}_{\text{out}})$ to bipartite quantum states on the respective spaces, $\mathscr{B}(\mathcal{H}_{\text{out}})\otimes \mathscr{B}(\mathcal{H}_{\text{in}})$.
\begin{equation}
\label{eq:cpmap}
    \mathcal{E}[\rho_{\text{in}}] = \text{tr}_E [u \ \rho_{\text{in}} \otimes \rho^{E} u^\dag] = \rho_{\text{out}}.
\end{equation}
In this context, the non-unitary evolution between the input and output states arises from a unitary interaction with an environment $E$. This type of channel is commonly represented using its Choi state through the projection
\begin{equation}
\label{eq:choi-act}
    \mathcal{E}[\rho_{\text{in}}] = \text{Tr}_{\text{in}}\left[(\mathbb{I}_{\text{out}}\otimes \rho_{\text{in}}^{\text{T}})\hat{\mathcal{E}}\right].
\end{equation}
Explicitly, the Choi state $\hat{\mathcal{E}}$ of a channel $\mathcal{E}$ on a $d-$dimensional system is formed by the action of $\mathcal{E}$ on one half of an unnormalised maximally entangled state $\ket{\Phi^+} = \sum_{i=1}^d \ket{ii}$:
\begin{equation}\label{eq:choicptp}
    \hat{\mathcal{E}} := (\mathcal{E}\otimes \mathcal{I})\left[|\Phi^+\rangle\langle\Phi^+|\right] = \sum_{i,j=1}^d \mathcal{E}\left[|i\rangle\langle j|\right]\otimes |i\rangle\langle j|.
\end{equation}
Here, $\mathcal{I}$ represents the identity map. This is illustrated in Figure~\ref{fig:choi_reps}a, along with its extension to the many-time case, which we discuss in the following subsection.\footnote{We denote the Choi state of a map $\mathcal{A}$ as $\hat{\mathcal{A}}$, except for the process tensor $\mathcal{T}$, whose Choi state is denoted by $\Upsilon$.}


In the lab, one can characterise the map $\mathcal{E}$ in Eq.~\eqref{eq:cpmap} by a suitable choice of preparations and measurements. However, one does not have access to the environment and the unitary quantum transformations in Eq.~\eqref{eq:choi-act}. Nevertheless, the Stinespring dilation theorem~\cite{milz21} guarantees that for every completely positive map, there exists a system-environment unitary transformation of the form of the right-hand side of the same equation. Note that this is only under the condition that initial system-environment correlations are not present~\cite{Pechukas1994}.

\subsubsection{Instruments}
\label{subsec:inst}

Quantum channels are often used to describe the non-unitary or noisy dynamics of a system. However, they can also describe generalised measurement of a system. Let us reconsider Eq.~\eqref{eq:cpmap}, but we now replace the environment $E$ with an ancilla $A$ that we can measure:
\begin{equation}
\label{eq:instrment}
    \mathcal{A}_x[\rho_{\text{in}}] = _A\! \left< x \right| u \ \rho_{\text{in}} \otimes \rho^{A} u^\dag \left| x \right>_A = \ \rho_{\text{out}|x}.
\end{equation}

Above, an interaction between the system and the ancilla is engineered. Next, the ancilla is measured, and the outcome is $x$. This corresponds to the CPTP quantum channel
\begin{equation}
    \label{eq:instch}
    \hat{\mathcal{A}} = \sum_x \ket{x}\!\bra{x}_A \otimes \hat{\mathcal{A}}_x,
\end{equation}
whose input is a state of $S$ and its output is a classical flag on $A$ and the corresponding state of $S$. This generalises quantum measurements to include a well-defined post-measurement state~\cite{PhysRevA.80.022339}. A crucial difference between a quantum channel and the elements of an instrument $\{\mathcal{A}_x\}$ is trace preservation (TP). Quantum channels are CPTP, while elements of an instrument are CPTNI, which stands for completely positive and trace non-increasing. 

Just like a measurement, an instrument is said to be \textit{informationally complete} (IC) when $\{\mathcal{A}_x\}$ span the space $S$. Finally, a multitime instrument is often referred to as a tester. Both instruments and testers are endowed with the Choi form and other forms employed to describe quantum channels.

\begin{figure*}[t]
    \centering
    \includegraphics[width=\linewidth]{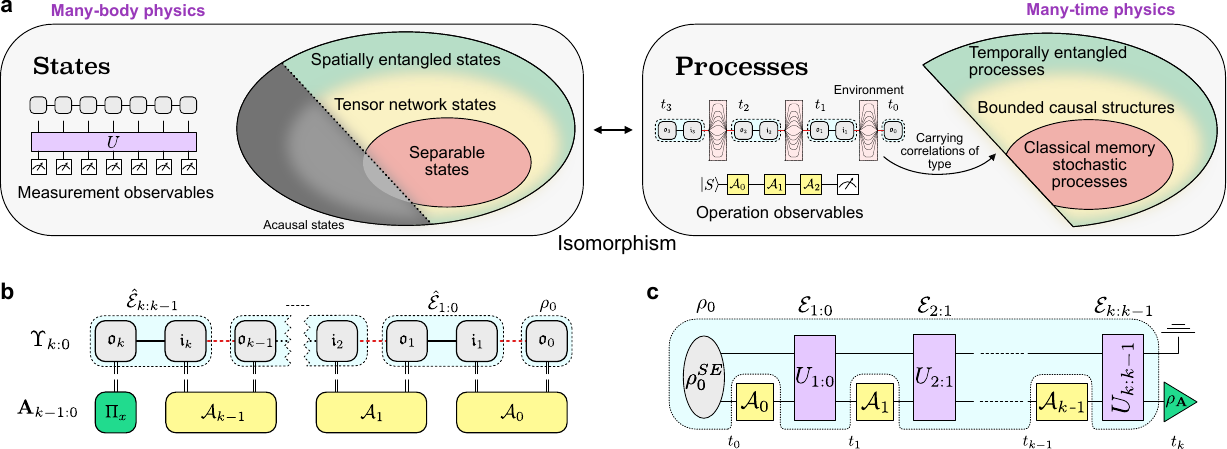}
    \caption{
    \textbf{a} 
    We refer to the Choi states of quantum processes as many-time states, which are isomorphic to many-body states modulo causality conditions. Hence, we can employ many of the tools developed to study many-body states to study quantum stochastic processes.
    \textbf{b} A many-time non-Markovian process is represented by a process tensor, which is a multitime density matrix that contracts with a sequence of instruments to yield a final state for the system.   
    \textbf{c} A $k$-step process tensor can also be seen as a sequence of correlated CPTP maps between different points in time, plus the effect due to the initial state. Output legs $\mathfrak{o}_l$ are mapped by a control operation $\mathcal{A}_l$ to the next input leg $\mathfrak{i}_{l+1}$.
    }
    \label{fig:subsystems_B}
\end{figure*}

\subsection{Process tensors}\label{sec:PT}
Quantum channels capture two-time correlations between the input and output states. As mentioned earlier, a typical quantum experiment may involve a sequence of control operations applied to the system at times $\mathbf{T}_k := {t_0, \cdots, t_k}$. This process is represented as a multilinear map that takes a sequence of controllable operations on the system and maps them to a final state density matrix. The map, referred to as a process tensor~\cite{Pollock2018a, 1367-2630-18-6-063032}, encapsulates all of the uncontrollable dynamics of the process., 

Process tensors formally generalize quantum channels to many-time processes. Specifically, we are interested in many-time quantum correlations that extend the two-time correlations captured by quantum channels. To be precise, we consider a situation where a $k$-step process is driven by a sequence $\mathbf{A}_{k-1:0}$ of control operations, each represented mathematically by CP maps that belong to an instrument: $\mathbf{A}_{k-1:0} := {\mathcal{A}_0, \mathcal{A}_1, \cdots, \mathcal{A}_{k-1}}$. After applying these operations, we obtain a final state $\rho_k(\mathbf{A}_{k-1:0})$, conditioned on the sequence of interventions. These controlled dynamics take the form:
\begin{equation}\label{eq:multiproc}
        \rho_k\left(\textbf{A}_{k-1:0}\right) = \text{tr}_E [U_{k:k-1} \, \mathcal{A}_{k-1} \cdots \, U_{1:0} \, \mathcal{A}_{0} (\rho^{SE}_0)],
\end{equation}
where $U_{k:k-1}(\cdot) = u_{k:k-1} (\cdot) u_{k:k-1}^\dag$ and $\mathcal{A}_{j}$ is the CP map applied at time $j$. Eq.~\eqref{eq:multiproc} is the many-time generalization of Eq.~\eqref{eq:cpmap}, where $\rho_{\text{in}}$ is replaced by a sequence of control operations $\mathbf{A}_{k-1:0}$. The process tensor $\mathcal{T}_{k:0}$ is defined as the mapping from past controls $\mathbf{A}_{k-1:0}$ to future states $\rho_k \left( \mathbf{A}_{k-1:0} \right)$:
\begin{equation}
\label{eq:PT}
\mathcal{T}_{k:0}\left[\mathbf{A}_{k-1:0}\right] = \rho_k(\mathbf{A}_{k-1:0}).
\end{equation}

It is often convenient to work with the Choi state of the process tensor. A generalization of the Choi-Jamiołkowski Isomorphism (CJI) allows this mapping to be represented as a many-body quantum state. The circuit representation of this CJI is shown in Figure~\ref{fig:choi_reps}b, alongside the standard CJI for quantum channels. The Choi states of multi-time processes possess the same features as the states of many-body systems. Thus, we have the temporal parallel with many-time observables, many-time states (be they separable, discordant, or entangled), temporal phases, etc. We dub it as \textit{many-time physics}, where these notions have an equivalent meaning in processes. We depict this in Figure~\ref{fig:subsystems_B}a. In particular, we will apply the recently discovered tool called \textit{classical shadow tomography} to the many-time setting.

To compute the action of the process tensor on any controlled sequence of operations, in terms of the Choi states of the process and the control operation translates to
\begin{equation}\label{eq:PToutputF}
    \rho_k(\mathbf{A}_{k-1:0}) \!=\! \text{Tr}_{\overline{\mathfrak{o}}_k} \! \left[ \left(\mathbb{I}_{\mathfrak{o}_k}\otimes \hat{\mathcal{A}}_{k-1}\otimes \cdots \otimes \hat{\mathcal{A}}_0 \right)^\text{T} \Upsilon_{k:0} \right].
\end{equation}
Above, $\overline{\mathfrak{o}}_k$ represents all indices except for $\mathfrak{o}_k$, as shown in Figure~\ref{fig:subsystems_B}b. This equation includes all intermediate system-environment ($SE$) dynamics as well as any initial correlations, demonstrating how sequences of operations form observables of the process tensor. Finally, the process tensor itself is depicted in Figure~\ref{fig:subsystems_B}c.

We emphasise our conventions on time throughout this work: in circuit diagrams, time runs from left to right in accordance with the literature. In matrix and tensor representations, however, time runs from right to left, in accordance with the standards of matrix multiplication. Although the tension between these two is unfortunate, we believe it simplifies matters in practice.


\subsection{Properties of quantum processes}\label{sec:props}

The Choi state is an operator on multipartite Hilbert spaces as
\begin{equation}\label{eq:processtensor}
    \Upsilon_{k:0} \in \mathcal{B}(\mathcal{H}_{\mathfrak{o}_k} \otimes \mathcal{H}_{\mathfrak{i}_{k-1}} \otimes \mathcal{H}_{\mathfrak{o}_{k-2}} \otimes \ldots \otimes \mathcal{H}_{\mathfrak{i}_{1}} \otimes \mathcal{H}_{\mathfrak{o}_0}).
\end{equation}
Each time step is associated with an output ($\mathfrak{o}$) and input ($\mathfrak{i}$) leg of the process. If the dynamical process is non-Markovian, the system-environment interactions will distribute temporal correlations as spatial correlations between different legs of the process tensor. These correlations can then be probed using a variety of established quantum or classical many-body tools. In Figure~\ref{fig:choi_reps}b, we have depicted these correlations in red. In such cases, the process tensor can be expanded in terms of a family of CP maps as:
\begin{equation}\label{eq:nmpt}
    \Upsilon_{k:0}^{(\text{non-Markov})} = \sum_{\mu} \alpha_\mu \ \hat{\mathcal{E}}_{k:k-1}^{\mu}\otimes \hat{\mathcal{E}}_{k-1:k-2}^{\mu}\otimes\cdots\otimes \hat{\mathcal{E}}_{1:0}^{\mu}\otimes\rho_{0}^{\mu},
\end{equation}
where $\{\hat{\mathcal{E}}_{y:x}^\mu\}$are CP maps from time $x$ to $y$. The index $\mu$ and the amplitudes $\alpha_\mu$ account for correlations in time. This is depicted in Figure~\ref{fig:subsystems_B}b. For classical non-Markovian processes, each $\hat{\mathcal{E}}^\mu$ is a CPTP channel and $\alpha_\mu \ge 0$.

In contrast, the correlations between consecutive $\mathfrak{i}/\mathfrak{o}$ legs (presented in black) are expected to be very high for nearly unitary processes and do not constitute a measure of non-Markovianity. In this vein, it can be shown that a process is Markovian if and only if it can be expressed in the form~\cite{Pollock2018}:
\begin{equation}
\Upsilon_{k:0}^{(\text{Markov})} = \hat{\mathcal{E}}_{k:k-1}\otimes \hat{\mathcal{E}}_{k-1:k-2}\otimes\cdots\otimes \hat{\mathcal{E}}_{1:0}\otimes\rho_0,
\end{equation}
where all $\hat{\mathcal{E}}$ are required to be CPTP. However, to operationally detect non-Markovian correlations in a process, we need to test if the future process is correlated with the past process. This can be done using a \textit{causal break}, i.e., looking at inequalities of the following form
\begin{equation}\label{eq:NMC}
\frac{\mbox{Tr}[(\hat{\mathbf{A}}^{(y)}_{\mathfrak{o}_{k} \cdots \mathfrak{i}_{j+1} } \otimes \hat{\mathbf{A}}^{(x)}_{\mathfrak{o}_j \cdots \mathfrak{o}_{0}} )^\text{T} \Upsilon_{k:0}]}{\mbox{Tr}[(\hat{\mathbf{A}}^{(x)}_{\mathfrak{o}_j : \mathfrak{o}_{0}} )^\text{T} \Upsilon_{k:0}]}
\!\ne\!
\frac{\mbox{Tr}[(\hat{\mathbf{A}}^{(y)}_{\mathfrak{o}_{k} \cdots \mathfrak{i}_{j+1} } \otimes \hat{\mathbf{A}}^{(x')}_{\mathfrak{o}_j \cdots \mathfrak{o}_{0}} )^\text{T} \Upsilon_{k:0}]}{\mbox{Tr}[(\hat{\mathbf{A}}^{(x')}_{\mathfrak{o}_j : \mathfrak{o}_{0}} )^\text{T} \Upsilon_{k:0}]}.
\end{equation}
The above equation asks whether the probabilities of observing an outcome $y$ in the future $(\mathfrak{o}_{k} \cdots \mathfrak{i}_{j+1})$ process differ, given that we observed two distinct outcomes $x$ and $x'$ in the past at $(\mathfrak{o}_{j} \cdots \mathfrak{o}_{0})$? The denominators are the probability of observing $x,x'$, which we do not care about. 

Importantly, we must break up the instrument at $j$ into a past and a future component -- across $\mathfrak{o}_j$ and $\mathfrak{i}_{j+1}$ -- this is the aforementioned causal break. It ensures that the future depends on the past if and only if the process is Markovian. This is because, when the instrument at $j$ is broken no information passes from the past to the future via the system.\footnote{This condition can be made stronger by multi-time conditionings, i.e. we may think of $y,x,x'$ to be a sequence of outcomes.} Note, that to test the above inequality, at least two measurements must be made in a single experiment, one at $\mathfrak{o}_k$ and another at $\mathfrak{o}_j$. The above test for non-Markovianity is not possible without a trace non-increasing instrument. Such quantum operations have Choi states with non-trivial local expectation values. In particular, the respective characteristic features of these are:
\begin{equation}\label{eq:nutd}
\hat{\mathcal{A}}_{\text{NU}}[\mathbb{I}] \neq \mathbb{I} \qquad  \mbox{and} \qquad
\text{Tr}_{\text{out}}[\hat{\mathcal{A}}_{\text{TD}}] \prec \mathbb{I}.
\end{equation}
That is, non-unitals map the maximally mixed state elsewhere, and trace-decreasing maps are non-deterministic.

The generation of temporal correlations across multiple times has a different physical origin. These correlations emerge when the system interacts strongly with a complex environment. Information can travel into a (quantum or classical) bath and later return to interact with the system, redistributing correlations~\cite{Milz2021PRXQ, rivas-NM-review}. The properties of this non-Markovian memory depend on both the system-environment interaction and the coherence of the bath. Such non-Markovian quantum stochastic processes can be fully described within the process tensor framework~\cite{Pollock2018a, Milz2021PRXQ}. For more general processes, such as those involving indefinite causal orders, the process matrix formalism can be used~\cite{Shrapnel_2018}. This broader framework allows for the description of more complex dynamical scenarios, where the causal structure itself may not be fixed or well-defined at all times.

It is important to note that although the set of process tensors is isomorphic to a subset of quantum states, processes are much more restricted in terms of the structure they can possess due to causality constraints. Causality appears in the form of a containment property in process tensors, such that averaging over measurements on an output leg produces the exact process up until that point:
$\text{Tr}_{\mathfrak{o}_k}[\Upsilon_{k:0}] = \Upsilon_{k-1:0}\otimes \mathbb{I}_{\mathfrak{i}_j}$. 
In other words, future operations do not affect past statistics. This provides a time ordering to the structure of correlations in processes. One example of this is that rank-one processes must be purely Markovian\footnote{Assuming we are not given the full environment at the end~\cite{PRXQuantum.5.010314}.} -- i.e., any many-time structure must be mediated by the surrounding environment. The causal requirement and the mechanism by which correlations are accessed are the primary differences between states and processes, and the focus of our exploration in this work. As we will discuss, process observables are non-trivial to handle and have different structures compared to state observables. For example, they can be deterministic at different times.

\subsection{Process tensor tomography}
\label{methods:PTT}

The process tensor framework is designed to experimentally access many-time quantum correlations. Recently, we introduced process tensor tomography (PTT) as a method for estimating many-time correlations on real devices~\cite{White-NM-2020, White-MLPT}. We now aim to translate these theoretical tools into practical techniques for use in experiments.

In any experiment, a sequence of CP operations ${\mathcal{A}_0, \mathcal{A}_1, \cdots, \mathcal{A}_{k-1}}$ is applied to the system. The collective Choi state of the full sequence is denoted as $\hat{\mathbf{A}}_{k-1:0} = \bigotimes_{i=0}^{k-1} \hat{\mathcal{A}}_i$. The tensor product naturally arises because these operations can be chosen independently. After applying these operations, the final state of the system is measured with a set of measurement operators ${M_x}$. The output distribution, conditioned on these operations and the measurement apparatus, is given by the spatiotemporal generalization of Born's rule~\cite{Shrapnel_2018}:
\begin{equation}
    \label{eq:PT_action}
    p_{x|\mathbf{A}_{k-1:0}} = \text{Tr}\left[\Upsilon_{k:0}(M_x \otimes  \hat{\mathbf{A}}_{k-1:0}^{\text{T}})\right].
\end{equation}

This generalizes the joint probability distributions of a classical stochastic process to the quantum regime~\cite{Milz2020}. The key insight of Eq.~\eqref{eq:PT_action} is that operations on a system at different points in time form many-time observables on the process tensor Choi state. These operations can be applied deterministically, such as with a unitary operation\cite{MTP1, PT-limited-control}, or stochastically, such as through measurements with feed-forward or more general quantum instruments. In this paper, we focus on the latter case; informationally complete instruments are essential for full tomography and for measuring the non-Markovianity of a process. It is important to note that many-time processes suffer from the same dimensionality curse as states and multipartite classical distributions. Specifically, as the number of timesteps increases, the number of histories that must be accounted for grows exponentially. This presents a severe limitation, which we address in our approaches.

A first step in helping overcome the curse of large spaces is to use maximum-likelihood estimation (MLE) for process tensor tomography (PTT), as described in Ref.~\cite{White-MLPT}. This reduces the burden of high-dimensional statistical effects, and allows for a positive and causal representation--as we shall see later, this can be used to conduct efficient tomography. In this work, we apply this tool to both restricted and unrestricted process tensors. We now provide a brief overview of MLE-PTT.

To fully characterize the process tensor, we must apply an informationally complete instrument ${\mathcal{B}j^{\mu_j}}$ at each time step. Next, we estimate the joint probability for the sequence of outcomes, denoted by $p_{i, \vec{\mu}}$. These probabilities are then used in a set of linear equations, which uniquely determine $\Upsilon_{k:0}$. Specifically, applying Eq.~\eqref{eq:PT_action} to the basis elements yields:
\begin{equation}
    p_{i,\vec{\mu}} = \text{Tr}\left[ \Upsilon_{k:0} (\Pi_i\otimes \hat{\mathcal{B}}_{k-1}^{\mu_{k-1}\text{T}}\otimes \cdots \otimes \hat{\mathcal{B}}_0^{\mu_0\text{T}}) \right].
\end{equation}
The final instrument becomes a POVM (positive operator-valued measure) because we do not require the post-measurement state at that stage.\footnote{If the probabilities are known only for a subset of operations.}

In practice, however, the measured outcomes $n_{i,\vec{\mu}}$ are noisy estimates of the ``true" probabilities, and it is often the case that no physical process tensor perfectly matches the data. For this reason—consistent with common practice—we treat the estimation of a process as an optimization problem, where a model for the process is fitted to the data. Specifically, a unique $\Upsilon_{k:0}$ is determined by minimizing the log-likelihood function:
\begin{equation}
    f(\Upsilon_{k:0}) = \sum_{i,\vec{\mu}}-n_{i,\vec{\mu}}\ln p_{i,\vec{\mu}},
\end{equation}
where $\Upsilon_{k:0}$ is a positive matrix that satisfies a set of causality conditions. This optimization is performed using a projected gradient descent algorithm, where at each step the model is updated in a direction that reduces the log-likelihood while ensuring that the updated model remains constrained to the manifold of positive, causal states. For more details on the process, refer to Ref.~\cite{White-MLPT}.

\section{Spatiotemporal Classical Shadows}
\label{sec:shadow}

In this section, we will extend the state-of-the-art in learning properties of quantum states to quantum stochastic processes. To do so, we will make use of the properties of quantum stochastic processes discussed in the last section, from both theoretical and practical perspectives. This extension will enable learning the properties of non-Markovian quantum stochastic processes, even when dealing with large numbers of steps or qubits.

Recent seminal work by Huang et al.~\cite{Huang2021,huang-shadow} introduced the concept of classical shadows, a technique that uses randomised measurements and post-processing to estimate $M$ quantum state observables in $\log M$ measurements, with additional scaling factors depending on the properties of the observable and the randomisation procedure~\cite{huang-shadow}. This concept has been expanded to include quantum process tomography and the estimation of gate set properties~\cite{Kunjummen2021,Helsen2021}. In this section, we demonstrate the natural application of classical shadows to multi-time, multi-qubit processes and address causally-imposed idiosyncrasies. The issue of causality highlights a possible learnability gap between processes and states, particularly in physically plausible scenarios. That is, causal ordering typically implies that correlations are high weight. But classical shadows is mostly only able to capture low-weight properties. We employ this method to investigate various properties of quantum stochastic processes and identify potential applications that can overcome the limitations imposed by the inherent structure of these processes.

By virtue of the state-process equivalence for multi-time processes~\cite{chiribella_memory_2008,Shrapnel_2018,Pollock2018a,PhysRevA.98.012328}, quantum operations on different parts of a system at different times constitute observables on a many-body quantum state. Consequently, properties of quantum stochastic processes can be similarly probed using state-of-the-art quantum state characterisation techniques. Classical shadow tomography~\cite{huang-shadow, elben2022randomized} is one such technique and already has many generalisations and applications~\cite{helsen2021estimating, hadfield2022measurements, huang2022provably, PhysRevLett.125.200501}. Measuring classical shadows allows for exponentially greater observables to be determined about a state, provided that those observables are sufficiently low-weight. But this restriction
means the technique has limitations for the study of temporal correlations in contrast to spatial ones, as we shall see. Namely, typical temporal correlations appear in high-weight observables. Hence, the full set of low-weight correlations in principle detectable by classical shadows is shrunk considerably by conditions of causality on the process. The free operation in classical shadows -- the identity observable -- is a maximally depolarising channel in our context. At a minimum, this limits detections of past-future correlations to those generated by non-unital dynamics, and partially scrambles the environment signal. 

We first derive the extension of classical shadows applied to process tensors. Specifically, we show how one can achieve the same exponential improvement in determining spatiotemporal properties of process tensors through randomised measurements. We perform this first in the idealised case of perfect Clifford measurements and state preparations, and then in the more general case of noisy quantum instrument bases, such as those we will encounter in Sec.~\ref{sec:IC-control}. There are many surprising limitations of classical shadows when applied to processes that we raise along the way. 
In subsequent sections, we then look to some applications of classical shadows in the process setting to determine key features of quantum stochastic processes.

\subsection{A Brief Overview of Classical Shadows}

We start with reviewing the key results on classical shadows whose details can be found in Refs.~\cite{huang-shadow,elben2022randomized,Kunjummen2021}. Given a fixed $n-$qubit state $\rho$, one sets a fixed ensemble of unitary rotations from which the state is rotated $\rho\mapsto U\rho U^\dagger$, and then projectively measured to generate a length $n$ bit string $|\tilde{b}\rangle \in \{0,1\}^n$. This series of steps generates a snapshot of the state which can be stored efficiently in classical memory: $U^\dagger |\tilde{b}\rangle\!\langle \tilde{b}|U$. 
Define a quantum channel $\mathcal{M}$ which is the averaging of both classically random unitary and quantumly random measurement outcome 
\begin{equation}
	\label{eq:shadow-channel}
	\mathcal{M}(\rho) = \mathbb{E}\left[U^\dagger |\tilde{b}\rangle\!\langle \tilde{b}| U\right]\implies \rho = \mathbb{E}\left[\mathcal{M}^{-1}\left(U^\dagger |\tilde{b}\rangle\!\langle \tilde{b}| U\right)\right].
\end{equation}
$\mathcal{M}^{-1}$ is not {CP}, but can still be applied to the measurement outcomes $U^\dagger|\tilde{b}\rangle\!\langle\tilde{b}|U$ in post-processing. Note that $\mathcal{M}$ is invertible for a tomographically complete set of measurements, but later we discuss tomographically incomplete cases where we invert only on the support of $\mathcal{M}$. A single classical snapshot of the state $\rho$ is given by $\tilde{\rho} = \mathcal{M}^{-1} \left(U^\dagger |\tilde{b}\rangle\!\langle \tilde{b}|U\right)$. In accordance with Eq.~\eqref{eq:shadow-channel}, this snapshot reproduces the state in expectation. Repeating the procedure $N$ times results in a collection of $N$ independent, classical snapshots of $\rho$:
\begin{equation}
	\label{eq:shadow-def}
	\mathtt{S}(\rho;N) = \left\{\tilde{\rho}_1 = \mathcal{M}^{-1}\left(U_1^\dagger|\tilde{b}_1\rangle\!\langle\tilde{b}_1|U_1\right), \cdots, \tilde{\rho}_N = \mathcal{M}^{-1}\left(U_N^\dagger|\tilde{b}_N\rangle\!\langle\tilde{b}_N|U_N\right)\right\}.
\end{equation}
This array is called a \emph{classical shadow} of $\rho$. We will now state, but not show, some results on classical shadows. 

A sufficiently sized classical shadow is expressive enough to predict many different observables of the state. For an $N$ shot classical shadow, the procedure involves computing observables on each snapshot and averaging the data. Because the snapshots are efficiently storable, this processing is also efficient. In practice, to avoid corruption from outliers, the snapshots are chunked into $K$ different sets, the observables computed on those sets, and the median value taken (so-called median-of-means algorithm).
To estimate a set of $M$ observables $\{O_1,\cdots,O_M\}$ from a classical shadow $\mathtt{S}(\rho,N)$, average the $K$ groups of size $\lfloor N/K \rfloor$ 
\begin{equation}
	\tilde{\rho}_{(k)} = \frac{1}{\lfloor N/K\rfloor}\sum_{i= (k-1\lfloor N/K \rfloor + 1)}^{k\lfloor N/K\rfloor}\tilde{\rho}_i,
\end{equation}
Then compute 
\begin{equation}
	\tilde{o}_i(N,K) := \text{median}\{\Tr[O_i\tilde{\rho}_{(1)}],\cdots, \Tr[O_i\tilde{\rho}_{(k)}]\}.
\end{equation}
for each $O_i$. 
This procedure comes equipped with the following set of guarantees: for $M$ observables on an $n$-qubit system $\{O_i\}$ select accuracy parameters $\epsilon,\delta\in [0,1]$, set 
\begin{equation}
	K = 2\log(2M/\delta)\quad\text{and}\quad N = \frac{34}{\epsilon^2}\max_{1\leq i \leq M} \|O_i - \frac{\Tr[O_i]}{2^n}\mathbb{I}\|^2_{\text{shadow}},
\end{equation}
where $\|\cdot\|_{\text{shadow}}$ is an operator norm that depends on the measurement ensemble. Then a classical shadow with $NK$ snapshots accurately determines the observables through the median-of-means estimation 
\begin{equation}
	|\tilde{o}_i(N,K) - \Tr[O_i\rho]|\leq \epsilon\:\forall \:1\leq i \leq M
\end{equation}
with probability at least $1-\delta$. 

We can see from this that the protocol does not a priori depend on the properties under scrutiny. This flexibility is one aspect to the appeal of shadows: the measurement primitive is independent of the scientific question, and hence a wide range of fundamentally different questions can be asked about a quantum state through the same set of data. This spans the estimation of local observables, entanglement verification, entanglement detection, computation of Renyi entropies, and can be generalised much further. We will discuss the effects of the operator norm in a later section. The protocol most used involves sampling each unitary from the set of local Cliffords and then projectively measured. The single-qubit inverted channel $\mathcal{M}_1^{-1}(X) = 3X - \mathbb{I}$, and across $n$ qubits scales to $\mathcal{M}_{\mathcal{C}_n}^{-1} = \bigotimes_{i=1}^n\mathcal{M}_1^{-1}$. The operator norm for this choice of measurements scales as $4^k\|O\|_{\infty}^2$, for which $k$ is the locality of the observable -- the number of qubits on which the observable acts non-trivially. In total, $\mathcal{O}(\log(M)4^k/\epsilon^2)$ local Clifford measurements are required to determine $M$ observables with locality at most $k$. Note that this result extends to the estimation of input-output pairs in a quantum channel~\cite{PhysRevResearch.6.013029}, as it will here in the multi-time setting.

\subsection{Temporal Classical Shadows}
\label{sec:tempshadows}

We now build on the results presented in the last section by generalising classical shadows procedure from states to process tensors. This turns out to be somewhat subtle as the causality conditions play a nontrivial role. This is our first theoretical result, which we will implement on real quantum hardware in the coming sections.

\par 
At each time $t_j$, the process has an output leg $\mathfrak{o}_j$ (which is measured), and input leg $\mathfrak{i}_{j+1}$ (which is a state that feeds back into the process).
The two important properties that we stress are: (i) a sequence of operations constitutes an observable on the process tensor via Eq.~\eqref{eq:PT_action}, generating the connection to classical shadows, and (ii) a process tensor constitutes a collection of possibly correlated {CPTP} maps and hence may be marginalised in both time and space to yield the $j$th {CPTP} map describing the dynamics of the $i$th qubit $\hat{\mathcal{E}}^{(q_i)}_{j:j-1}$.

We now construct the shadows procedure for processes with respect to general ensembles of instruments. The effect of the choice of ensembles will be to change the shadow norm. One might select this ensemble to be random Pauli measurements and the fresh preparation of a stabiliser state at each time. For this, the shadow norm once again will scale like $4^k\|O\|_\infty$. However, we consider the general case to be more pertinent, since this will include the experimentally realisable bootstrapped basis that we shall develop in~Sec.~\ref{sec:IC-control}. For local operations, one will always incur an exponential overhead in the weight, with at most a constant cost for the use of an `unideal' set of instruments. 
In principle, one could additionally study more global ensembles where the controls themselves are multi-time (via quantum or classical memory -- so-called \emph{testers}), but this is beyond the scope of the present work.



In the tomographic representation of a process tensor (see Ref.~\cite{Pollock2018a}), given a basis of operations at the $i$th time $\{\mathcal{B}_i^{\mu_i}\}$ with Choi states $\{\hat{\mathcal{B}}_i^{\mu_i}\}$. There exists a \emph{dual} set $\mathcal{D} := \{\Delta_j^{\mu_j}\}$ such that the process tensor may be written
\begin{equation}
	\label{eq:choi-pt-full}
	\Upsilon_{k:0} = \sum_{i,\vec{\mu}} p_k^{i,\vec{\mu}}\omega_i\otimes
	\Delta_{k-1}^{\mu_{k-1}\ \text{T}}\otimes \cdots \otimes \Delta_1^{\mu_1\ \text{T}}\otimes \Delta_0^{\mu_0\ \text{T}}.
\end{equation}
When a linearly independent basis is chosen, the duals satisfy $\text{Tr}[\hat{\mathcal{B}}_i\Delta_j^\text{T}] = \delta_{ij}$. Similarly, we have expanded the final state $\rho_k^{\vec{\mu}}$ in terms of the dual operators $\omega_i$ of the probing {POVM} $\{\Pi_i\}$, i.e., $\rho_k^{\vec{\mu}} = \sum_{i} p_k^{i,\vec{\mu}} \omega_i$.
Here, $p^{i,\vec{\mu}}$ is the probability of the detector clicking with outcome $i$ subject to some sequence of operations $\mathbf{B}_{k-1:0}^{\vec{\mu}}$. Another way to say this is that Eq.~\eqref{eq:choi-pt-full} is the linear inversion estimate of $\Upsilon_{k:0}$, as described in Sec.~\ref{methods:PTT}. 

Let $\mathcal{M}_i' : \Delta_i^{\mu_i} \mapsto \hat{\mathcal{B}}_i^{\mu_i}$, and let $\mathcal{M}' = \bigotimes_i\mathcal{M}_i'$. We see then that 
\begin{equation}
	\begin{split}
		\Upsilon_{k:0} &= \sum_{i,\vec{\mu}} p_k^{i,\vec{\mu}}\omega_i\otimes
		\Delta_{k-1}^{\mu_{k-1}\ \text{T}}\otimes \cdots \otimes \Delta_1^{\mu_1\ \text{T}}\otimes \Delta^{\mu_0\ \text{T}},\\
		&= \sum_{i,\vec{\mu}} p_k^{i,\vec{\mu}}
		\mathcal{M}_k'^{-1}(\Pi_i)\otimes \mathcal{M}_{k-1}^{-1}(\hat{\mathcal{B}}_{k-1}^{\mu_{k-1}})'\otimes \cdots \otimes \mathcal{M}_{1}^{-1}(\hat{\mathcal{B}}_{1}^{\mu_{1}})'\otimes \mathcal{M}_{0}^{-1}(\hat{\mathcal{B}}_{0}^{\mu_{0}})',\\
		&= \mathbb{E}\left[\mathcal{M}'^{-1}\left(\Pi_k\otimes \hat{\mathbf{B}}_{k-1:0}\right)\right].
	\end{split}
\end{equation}
We identify $\mathcal{M}'$ as being exactly the same channel as used to define classical shadows in Eq.~\eqref{eq:shadow-channel}.
As a consequence, the single-shot version of Eq.~\eqref{eq:choi-pt-full} is exactly a classical shadow. This connection points to a nice interpretation of classical shadow tomography: it is effectively a Monte-Carlo sampling of the tomographic representation of a quantum state. 
This also implies that if the basis is incomplete -- for example, restricted to the set of unitaries -- that classical shadows are still valid, but the state produced in expectation will be a restricted process tensor rather than a full one. 
We can now use our tools for computing dual sets to the problem of generating classical shadows on process tensors from arbitrary instruments.

To compute the classical snapshot from a sequence of instruments, first, let $N$ be the size of the (possibly under or overcomplete) basis. We can write each of the Choi $\hat{\mathcal{B}}_i^{\mu_i}$ as a row vector (adopting a row-vectorised convention) $\langle\!\langle\hat{\mathcal{B}}_i^{\mu_i}|$ and from this construct a single matrix $\hat{\mathcal{M}}'_i$:
\begin{equation}
	\hat{\mathcal{M}}'_i = \begin{pmatrix}
		\langle\!\langle\hat{\mathcal{B}}_i^{0}| \\ 
		\langle\!\langle\hat{\mathcal{B}}_i^{1}| \\ 
		\vdots\\
		\langle\!\langle\hat{\mathcal{B}}_i^{N-1}| \\ 
	\end{pmatrix}.
\end{equation}
Now, the matrix $\mathcal{M}_i'^{\ +\dagger}$ (right-inverse, conjugate transposed) contains the dual matrices $\{\Delta_i^{\mu_i}\}$ to $\{\hat{\mathcal{B}}_i^{\mu_i}\}$ in its rows, which are then trace normalised. 

In our forthcoming experiments on noisy intermediate-scale quantum (NISQ) devices, we use the parametrised set of bootstrapped instruments, to be described in Sec.~\ref{sec:IC-control}. Our basis set consists of the instruments generated by a local Clifford gate on the system sandwiched between two short interactions with an ancilla, plus a projective measurement outcome on that ancilla. For each randomised sequence, a sequence of Cliffords is applied to the system (recalling that these are each sandwiched between fixed $SA$ interactions), and the outcome of the series of mid-circuit measurements on the ancilla qubit recorded, $x_kx_{k-1}\cdots x_0$. The duals to this basis of instruments are computed, and the classical shadow of the process tensor represented from the snapshots
\begin{equation}
	\label{eq:bootstrapped-shadow}
	\hat{\Upsilon}_{k:0}^{\vec{x},\vec{\mu}} := \omega_{x_k} \otimes \Delta_{k-1}^{\mu_{k-1}(x_{k-1})\text{T}}\otimes \cdots \otimes \Delta_1^{\mu_1 (x_1)\text{T}}\otimes \Delta_0^{\mu_0(x_0)\text{T}}.
\end{equation}
Specifically, one would construct an analogous set of snapshots $\mathtt{S}(\Upsilon_{k:0},N)$ of the above form.
This provides a computationally convenient method to compute the classical shadow for an arbitrary ensemble of quantum instruments, with efficient computation of local observables as per the classical shadow routine. 

The scaling of the shadows procedure depends on the shadow-norm as it is defined below for projective measurements:
\begin{equation}
	\|O\|_{\text{shadow}} = \max_{\sigma : \text{state}} \left(\mathbb{E}_{U\sim\mathcal{U}}\sum_{\tilde{b}\in\{0,1\}^n} \langle \tilde{b}|U\sigma U^\dagger |\tilde{b}\rangle\!\langle \tilde{b}|U \mathcal{M}^{-1}(O)U^\dagger |\tilde{b}\rangle^2\right)^{1/2}.
\end{equation}
The $\langle \tilde{b}|U\sigma U^\dagger |\tilde{b}\rangle$ part of the expression is the probability of obtaining outcomes $\tilde{b}$ from state $\sigma$ in the basis of $U$, and the $\langle \tilde{b}|U \mathcal{M}^{-1}(O)U^\dagger |\tilde{b}\rangle$ is the expectation value of the snapshot with respect to the resulting outcome. Equivalently, for a given basis of high rank instruments, one can numerically compute the quality of the ensemble by solving the {SDP} for the single step observable
\begin{equation}
	\|O\|_{\text{shadow} \ i} = \max_{\sigma :\text{state}}\left(\mathbb{E}_{\hat{\mathcal{B}}\sim\mathscr{B}_i} \Tr[\sigma \hat{B}^{\text{T}}] \Tr[\mathcal{M}_i'^{-1}(O)\hat{B}^\text{T}]\right)^{1/2},
\end{equation}
from which $\|O\|_{\text{shadow}} = \prod_{i=0}^k\|O\|_{\text{shadow}\ i}$. 

\subsection{Recovering Useful Observables and Process Marginals}

We have generalised classical shadows to learning properties of quantum stochastic processes.
It is important now to consider: what interesting information can we actually extract using this method? In
later sections, we consider several different applications thereof. Let us first discuss the constraint of requiring low-weight observables, and subsequently the application to finding process marginals. 

\subsubsection{Low Weight Observables}

For all locally drawn ensembles -- and many global ones -- the procedure of classical shadows scales with an operator norm that is exponential in the locality $\ell$ of the observables. That is, the number of qubits that some operator $O$ acts non-trivially on. For a single tensor product Pauli string, this is the number of traceless Pauli operators present. Consider observables of this form:
\begin{equation}
	O = P_{\mathfrak{o}_k}\otimes P_{\mathfrak{i}_{k}}\otimes\cdots \otimes P_{\mathfrak{o}_1}\otimes P_{\mathfrak{i}_1}\otimes P_{\mathfrak{o}_0}.
\end{equation}
The implication here is that for any feasible set of experiments, the desired operators to learn must be mostly $\mathbb{I}$, with a handful of $X$, $Y$, and $Z$. In the Choi picture, the operator $\mathbb{I}$ has an operational meaning as the physical instrument of either a partial trace (measure and discard) in the outputs $\mathfrak{o}_j$, or a preparation of a maximally mixed state in the inputs $\mathfrak{i}_j$. As an instrument across $\mathfrak{o}_{j-1}\mathfrak{i}_j$, it represents a completely depolarising channel. 
In quantum states, there is no a priori structure to the set of correlations between qubits. However, with processes, we have seen in Sec.~\ref{sec:background} the effects of causality constraints. These ensure that there is a causal direction. But this is problematic, because information must travel in and out of the system -- and so any discarding of the system by low-weight observables equally erases important information. One concrete example is in a two-step unital quantum process from which we wish to extract two-point correlation functions. Causality forbids any $\mathfrak{i}_1-\mathfrak{i}_2$, and $\mathfrak{i}_1-\mathfrak{o}_0$ or $\mathfrak{i}_2-\mathfrak{o}_1$ correlations; unitality forbids $\mathfrak{o}_2-\mathfrak{o}_1$, $\mathfrak{o}_1-\mathfrak{o}_0$, and $\mathfrak{o}_2-\mathfrak{o}_0$ correlations. This leaves $\mathfrak{i}_1-\mathfrak{o}_2$ as the only non-trivial two-point correlator of the process. 

The problem more generally stems from the fact that dynamics build up non-trivial entanglement between a system and its environment. If any past leg of the system is rendered incoherent, this incoherence will destroy the system-environment entanglement, preventing future temporal entanglement. 
This unique structure of processes means we have to be somewhat careful and imaginative when applying the framework of classical shadows to the temporal domain.

\subsubsection{Process Marginals}
\label{sec:process_marg}

We find that one particularly useful application of classical shadows to quantum stochastic processes is simultaneous determination of process tensor marginals, particularly contiguous ones. This allows us to determine a subset of correlations between different times. More usefully, though, we shall explore in Sec.~\ref{sec:dynamic-sampling} different classes of processes that may be reconstructed from their marginals. 
A $k-$step process marginal across $n$ qubits is fully determined by a set of $\mathcal{O}(2^{2kn})$ weight $kn$ Pauli observables. 
When marginalising across all but a handful of times or qubits, we will denote the remaining steps or registers by commas, i.e.
\begin{equation}
	\label{PT-marg}
	\begin{split}
		\hat{\mathcal{E}}^{(q_{i_0},q_{i_1})}_{j_0:j_0-1,j_1:j_1-1}\!&:=\! \text{Tr}_{ {\{\overline{q_{i_0}},\overline{q_{i_1}}\}}, \{\overline{t_{j_0}},\overline{t_{j_0-1}},\overline{t_{j_1}},\overline{t_{j_1-1}}\}}[\Upsilon_{k:0}];\\
		\Upsilon_{k:0}^{(q_i)} &= \text{Tr}_{\overline{q_i}}[\Upsilon_{k:0}],
	\end{split}
\end{equation}
where the overlines denote complement. For example, if $\mathbf{Q}:=\{q_0,q_1,\cdots,q_{n-1}\}$ were to denote all qubits on a device, then $\overline{q_i} = \mathbf{Q}\backslash\{q_i\}$. However, for the remainder of this work we operate only on a single qubit and omit this label. \par

Although collecting shadows suffices to estimate properties of the process marginals, they will not in general be able to reconstruct a physical estimate of the process marginals themselves. To this effect, we estimate enough observables to constitute informationally complete information about the process tensor and then employ {MLE}-{PTT} to process the data and obtain a physical estimate. The advantage of a physical (positive, causal) estimate is that we make use of information-theoretic tools that rely on the positivity of the state. \par

Lastly, it is important to ascribe and emphasise an appropriate physical significance to process marginals. In a multi-partite state $\rho_{ABC}$, the state marginal $\rho_{AC}$ (i.e., after performing a partial trace on $B$) is the state one obtains after projectively measuring $B$ and discarding the outcome. In a multi-partite channel $\mathcal{E}_{ABC}$, the \emph{channel} marginal $\mathcal{E}_{AC}$ is the resultant channel one would obtain by conditionally inserting a maximally mixed state on system $B$, and then projectively measuring $B$ and discarding the outcome. \emph{Process marginals} inherit this same interpretation, except that now the multi-party channel corresponds to different inputs and outputs on the same system at various times. That is, for example, the process marginal $\hat{\mathcal{E}}_{3:2;1:0}$ is the conditional process one would obtain by inserting a maximally mixed state at time $t_1$ and measuring/discarding at $t_2$. Although this might seem like a distant abstraction from physical settings, it essentially corresponds to being ignorant of the system for a given window of time. For example, the single-step marginal $\hat{\mathcal{E}}_j$ is the resulting channel one would obtain if they performed quantum process tomography only for the evolution from time $t_{j}$ to $t_{j+1}$.

In this section, we extended the method of classical shadow tomography to the spatiotemporal domain, enabling the efficient identification of properties for multi-time, multi-qubit processes. Next, we will demonstrate this procedure and its applications in subsequent sections. The randomisation of our current ensembles produces property estimates that are efficient when perturbed around the identity, obtaining expectation values such as $\langle X \mathbb{I}\mathbb{I}\cdots \mathbb{I} Y\rangle$. However, causality constraints suggest that, in physically realistic scenarios, most non-trivial observables are likely to be highly non-local ones, involving coherent control at the system level. As a result, the applications we present are not only useful but also tailored to accommodate these restrictions. An intriguing question to explore is whether there exist accessible ensembles that allow for sampling and obtaining perturbations around high-weight observables, for expectation values such as $\langle X \Phi^+\Phi^+\cdots\Phi^+Y\rangle$. This could enable more sensitive probing of long-range non-Markovian memory. Nevertheless, in the following sections, we will examine some applications of classical shadow tomography in the setting of quantum stochastic processes.

\section{Manufacturing Informationally Complete Quantum Instruments}
\label{sec:IC-control}
A central obstacle to complete non-Markovian process characterisation is the absence of an informationally complete set of controls in practice.\footnote{The instrument must allows for extra control that is needed for tomographically complete observation of a process tensor, namely non-unital and trace-decreasing maps.} This is because mid-circuit measurements are not usually available, and complete characterisation of processes requires control beyond unitary transformations. Previously, we showcased the many-time features that are accessible on devices that typically only have unitary control~\cite{MTP1}. While it is possible to bound a whole host of temporal correlations with only unitary control, here we design and implement quantum instruments that may be used to probe non-Markovian correlations in practice. Our goal is to develop techniques for mid-circuit measurements in this section. This will enable us to perform temporal classical shadow tomography in the next section. This also means that we can access the full process tensors for any small number of time steps.

Functional mid-circuit measurements on current devices require long implementation times, in the range of a few microseconds in superconducting devices or dozens on ion traps. Hence, such estimates cannot adequately characterise the dynamics during that particular window. We can circumvent this problem by introducing an ancilla qubit. The central idea is that {IC} control of a qubit may be realised through a two-qubit unitary control operation with the aid of an ancilla qubit, and a projective measurement. By the principle of deferred measurement, once the system has interacted with its ancilla, it does not matter when that ancilla is measured. The system can continue to participate in any $SE$ dynamics, and the relevant time scale becomes only the $SA$ unitary interaction and not the measurement on $A$. However, in practice, this level of control will not be ideal unless the following requirements are met: control of this style must be (i) short compared to the surrounding dynamics, and (ii) a well-chacacterisable quantum operation. Introducing the ancilla overcomes criteria (i). To overcome criteria (ii), we must develop a method to uncover the exact structure of the probe we are constructing. 

Below we will make use of the standard one-qubit quantum logic gates. Let $H$ denote the Hadamard gate, $F$ denote the phase gate, and $X, \ Y, \ Z$ denote the Pauli gates. Each of these is defined as:
\begin{equation}
    H=\tfrac{1}{\sqrt{2}}\begin{pmatrix} 1 & 1 \\ 1 & -1\end{pmatrix},\
    F=\begin{pmatrix} 1 & 0 \\ 0 & i\end{pmatrix},\
    X=\begin{pmatrix} 0 & 1 \\ 1 & 0\end{pmatrix},\
    Y=\begin{pmatrix} 0 & -i \\ i & 0\end{pmatrix},\
    Z=\begin{pmatrix} 1 & 0 \\ 0 & -1\end{pmatrix}.
\end{equation}

\subsection{Engineering Informationally-Complete Control}

The tools to probe process tensors are quantum instruments -- stochastic maps with a post-measurement state -- as described in Sec.~\ref{subsec:inst}. We show here how to manipulate an interaction with an ancilla qubit, creating an arbitrary well-characterised effective instrument. We will `bootstrapped' the tools to required for probing unknown $SE$ dynamics, i.e. to reconstruct a process tensor. Our goal is to engineer a set of \textit{completely positive, trace-non-increasing} (CPTNI) maps $\{\mathcal{A}_x\}_{x\in \mathcal{X}}$. Here, $x$ indexes a finite alphabet $\mathcal{X}$ such that the sum map $\sum_{x\in\mathcal{X}}\mathcal{A}_x$ is a CPTP map. The output of the instrument can be defined as a classical-quantum channel as described in Eq.~\eqref{eq:instch}:
\begin{equation}\label{eq:instact}
    \mathcal{A}(\rho) = \sum_{x\in\mathcal{X}} |x\rangle\!\langle x|_A\otimes \hat{\mathcal{A}}_x(\rho).
\end{equation}
An IC instrument for a qubit has 16 elements. In general, if $\{\mathcal{A}_x\}_{x\in \mathcal{X}}$ linearly span the whole space, then $\mathcal{A}$ is {IC}. In contrast, a unitary instrument, such as those implemented in Ref.~\cite{White-NM-2020, MTP1} has only 10 elements~\cite{PT-limited-control}.

To start, we prepare the ancilla in the state $\ket{i+} := FH\ket{0}=  \tfrac{1}{\sqrt{2}}(\ket{0}+i\ket{1})$. Let the initial $S$ state be $\rho_S$ and the $SA$ interaction be 
\begin{equation}\label{eq:instpulse}
u_w = V_\gamma \  F \otimes w(\theta, \phi, \lambda)  \ V_\gamma \qquad \text{with} \qquad V_\gamma:=\text{exp}\left(-\frac{i\gamma}{2} X_A\otimes Z_S \right),
\end{equation}
where $V_\gamma$ is a cross-resonance $SA$ interaction~\cite{malekakhlagh2020first} that leads to a short non-Markovian process as per Eq.~\eqref{eq:instrment}. This is the native cross-resonance interaction on IBM Quantum devices -- for some $\gamma$.\footnote{The cross-resonance interactions are likely introduce extraneous local terms, so what results will not be pure $XZ$ but will be characterised nonetheless.} Next, a phase gate $F$ is then applied on $A$ and a unitary $w$ on $S$ parametrised as
\begin{equation}\label{eq:local}
    w(\theta,\phi,\lambda) = \begin{pmatrix}
    \cos(\theta/2) & -\text{e}^{i\lambda}\sin(\theta/2) \\
    \text{e}^{i\phi}\sin(\theta/2) & \text{e}^{i\lambda + i\phi}\cos(\theta/2)
    \end{pmatrix}.
\end{equation}
This is followed by another cross-resonance $SA$ interaction. 

Following the interaction $u_w$, the ancilla is projectively measured to observe outcome $x$. This gives us a two-step process tensor that is a mapping from two inputs on $S$, the initial state and $w$, to the final $SA$ state:
\begin{equation}
\begin{split}\label{eq:instPT}
    \Upsilon_{\text{inst}} (\mathcal{W}(\theta,\phi,\lambda), \rho_S) :=&
    \sum_x \ket{x}\!\bra{x}_A
    \left( u_w \ket{i+}\!\bra{i+}_A \otimes  \rho_S \ u_w^\dag \right) \ket{x}\!\bra{x}_A \\
    =& \ket{0}_A\!\bra{0}\otimes \rho_{S|0}'+ \ket{1}_A\!\bra{1}\otimes \rho_{S|1}'.
\end{split}    
\end{equation}
Above, $\mathcal{W} (\cdot)= w(\theta,\phi,\lambda) (\cdot) w^\dag (\theta,\phi,\lambda)$ and $\rho_S$. The right-hand side of the last equation defines the instrument as given in Eq.~\eqref{eq:instact}. This means we can write down the corresponding instrument as
\begin{equation}
	\label{eq:PTT-instrument}
	\hat{\mathcal{A}} (\theta,\phi,\lambda) = \sum_x
\ket{x}\!\bra{x}_A \otimes
\sum_{i=0}^3
\bra{x}
\Upsilon_{\text{inst}} (\mathcal{W}(\theta,\phi,\lambda), \rho_i)
\ket{x}
 \otimes \omega_i^{\text{T}},
\end{equation}
where the middle term is output state $\rho_{i|x}'$ of the system and $\omega_i^\text{T}$ is the dual matrix to a basis of input density matrices $\rho_i$ such that $\text{Tr}[\rho_i \omega_j^\text{T}] = \delta_{ij}$. 

\begin{figure}[h!]
	\includegraphics[width=\linewidth]{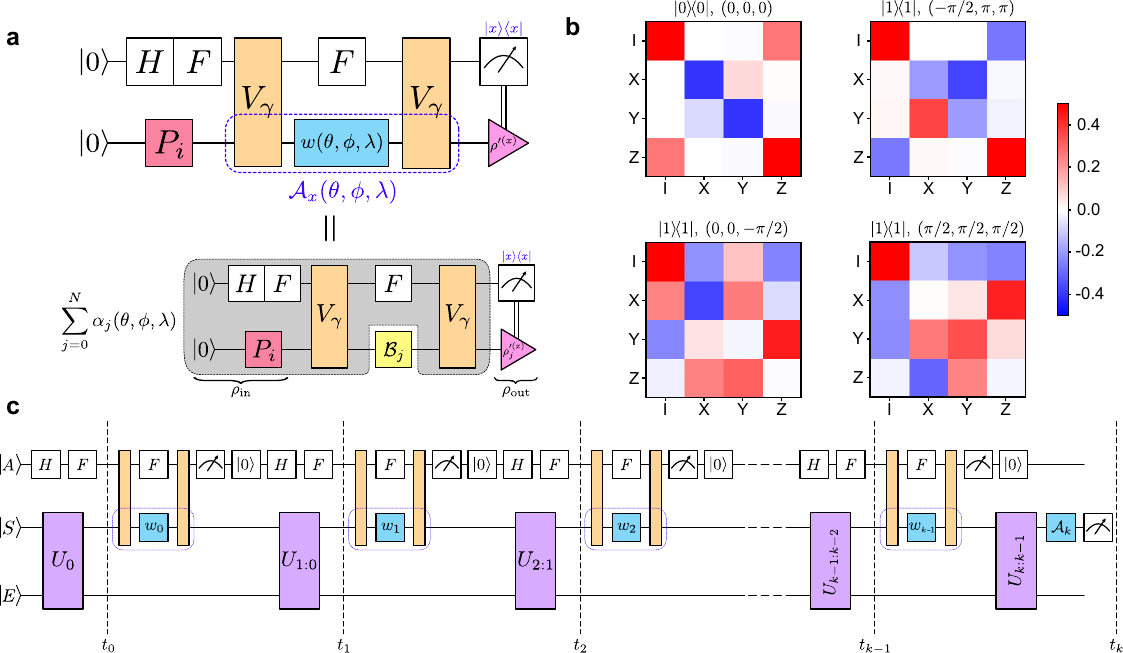}
	\caption[Instrument design for complete reconstruction of multi-time statistics ]{Instrument design for complete multi-time statistics. \textbf{a} Using a structured system-ancilla interaction, we construct a restricted process tensor with local unitaries to realise a locally-controllable instrument. This gives an exact characterisation of the effective instrument $\mathcal{A}_x(\theta,\phi,\lambda)$ for any choice of parameter values. Importantly, as per Eq.~\eqref{eq:PTM-vals}, these instruments will access the full trace-decreasing and non-unital subspace of superoperator space. \textbf{b} Some sample basis instruments implemented using this method on \emph{ibm\_perth} written in the Pauli basis, showing {IC} controllability including non-unital and trace-decreasing directions. These PTMs demonstrate that well-conditioned {IC} instrument sets can be achieved.\textbf{c} An illustration of how these instrument sets can be used to probe multi-time statistics. Each multi-time instrument is realised with the circuit as shown -- by applying the locally-modulated interaction with the ancilla, measuring (and recording) the ancilla, and subsequently resetting it. The different instruments applied capture temporal quantum correlations from an arbitrarily strong $SE$ interaction.}
	\label{fig:instrument-creation}
\end{figure}

This scheme gives us a process tensor with a mapping from the local pulse parameters to the effective operation on the system. A circuit for this scheme is depicted in Figure~\ref{fig:instrument-creation}a. The resulting object is a quantum instrument on $S$, i.e., CPTNI maps $\mathcal{A}_x (\theta,\phi,\lambda)$, which are functions of the local unitary $w(\theta,\phi,\lambda)$ and the $SA$ interaction. This scheme is designed to make all directions in this instrument space accessible. Namely, $\mathcal{A}_x(\theta,\phi,\lambda)$ should possess non-unital and trace-decreasing features, see Eq.~\eqref{eq:nutd}. We can find the set of parameters that lead to a desirable set of quantum instruments to amplify the six entanglement-breaking directions. This is most easily seen by examining the Pauli transfer matrix (PTM) $R_\mathcal{A}$ corresponding to a channel $\mathcal{A}$: $(R_{\mathcal{A}})_{ij} = \text{Tr}\left[P_i \mathcal{A}[P_j]\right]$, where $P \in {\mathbb{I}, X, Y, Z}$ are the Pauli operators.\footnote{In terms of the Choi state of the channel, the PTM is given as $(R_\mathcal{A})_{ij} = \text{Tr}\left[P_i \otimes P_j^{\text{T}}\hat{\mathcal{A}}\right],$ where Pauli $Y$ operator must be transposed.} The PTM elements corresponding to a $x=0$ measurement on the ancilla, with $\gamma=\pi/4$ are:
\begin{equation}
\label{eq:PTM-vals}
	\begin{split}
		(R_{\mathcal{A}_0})_{x0} &= \frac{\sqrt{2}}{2}\cos\phi\sin\theta - 2\cos\frac{\theta}{4}\sin^3\frac{\theta}{4}\sin\phi,\\
		(R_{\mathcal{A}_0})_{y0} &= \frac{1}{4}\cos\phi\left(2\sin\frac{\theta}{2} - \sin\theta\right) + \frac{\sqrt{2}}{2}\sin\theta\sin\phi,\\
		(R_{\mathcal{A}_0})_{z0} &= \frac{1}{8}\left(-1 + 5\cos\theta\right),\\
		(R_{\mathcal{A}_0})_{0x} &= \sin\frac{\theta}{2}\left(\sqrt{2}\cos\lambda\sin^2\frac{\theta}{4} + \cos^2\frac{\theta}{4}\sin\lambda\right),\\
		(R_{\mathcal{A}_0})_{0y} &= \frac{1}{2}\sin\frac{\theta}{2}\left[\left(1+\cos\frac{\theta}{2}\right)\cos\lambda - \sqrt{2}\left(1 - \cos\frac{\theta}{2}\right)\sin\lambda\right],\\
		(R_{\mathcal{A}_0})_{0z} &= \frac{1}{8}(3+\cos\theta).
	\end{split}
\end{equation}
The PTM is valuable for distinguishing three physically distinct vector spaces: those consisting of linear combinations of unitary operators, maps that ``un-mix'' or restore the identity operator, and stochastic maps that reduce the trace of a state. The first three elements quantify the size of non-unitality and the latter three quantify the size of trace non-increasingness. This demonstrates a high level of control, by varying $\{ \theta, \phi, \lambda\}$, to achieve desirable properties for the instrument.

\subsection{Characterising Informationally-Complete Control}


In practice, we fully characterise the restricted process tensor given in Eq.~\eqref{eq:instPT} using PTT from Sec~\ref{methods:PTT}. We do this because the instrument implementation in practice will be noisy: the cross-resonance interaction is imperfect and generates additional local terms in the Hamiltonian as well as slight decoherence. Moreover, the measurements themselves made on each qubit have some associated error. These issues may be well accounted for as long as the instrument is characterised. However, employing quantum process tomography (QPT) will only be valid for a single set of local parameter values -- and it would be highly expensive to repeat the procedure for every desired local gate. Instead, we perform restricted {PTT} with a basis of unitary operations on the system for an informationally complete set of measurements. This only requires 40 circuits.


For this we set $\rho_S = \ket{0}\!\bra{0}$ in Eq.~\eqref{eq:instPT} and prepare the initial state as $\rho_i := \mathcal{P}_i(\rho_S) := \pi_i (\cdot) \pi^\dag_i $ with $\pi_i = \{H, FH, \mathbb{I}, X\}$, which is an IC set of preparations:
\begin{equation}
\rho_i =\{\ket{+}\!\bra{+}, \ket{i+}\!\bra{i+}, \ket{0}\!\bra{0},\ket{1}\!\bra{1}\}.
\end{equation}
We then follow the procedure above and characterise the output states as a function of $w(\theta,\phi,\lambda)$ and the measurement on $A$.

For optimal characterisation, we first perform gate set tomography~\cite{gst-2013} to obtain estimates of both the system input states $\rho_i$ and the system and ancilla {POVM}s. This ensures that the resulting {PTT} to estimate $\Upsilon_{\text{inst}}$ and {QPT} on the instrument is self-consistent. Using Eq.~\eqref{eq:PTT-instrument} we then have an accurate representation of the effective instrument acting on the system for any choice of $\theta$, $\phi$, and $\lambda$. 
Figure~\ref{fig:instrument-creation}b shows some representative basis instruments that have been experimentally reconstructed after applying this procedure on \emph{ibm\_perth}.
Finally, noting that we may reset the ancilla qubit after each projective measurement to reuse it, this permits us the necessary control for tomographically complete observation of a multi-time process. Having demonstrated that these instruments permit us complete control of the system, we can use them for complete process characterisation as illustrated in Figure~\ref{fig:instrument-creation}c. For concreteness, these elements consist of various choices of the local $w$ and a two-outcome value $x$, from which can can determine the effective $\mathcal{A}_x$

In the case where mid-circuit measurement capabilities are not available, we can still use ancilla qubits as instruments, but the measurement needs to be deferred until the very end of the circuit. This has two limitations: the first is that there now needs to be one ancilla qubit for each instrument, they cannot be reused. Practically, then, this limits us to 2--3 steps. Secondly, the ancilla qubits themselves may interact with the system (influencing the process tensor), or they may thermalise, changing the measurement populations. This means that the processes cannot be on too long a timescale. We circumvent the interactions by applying dynamical decoupling protocols to the ancilla qubits. This type of process tensor is restricted in scope but does demonstrate that even devices without any mid-circuit measurements can still achieve tomographically complete control. Note that in principle a single ancilla qubit suffices to be used as a tester across an arbitrary number of steps but in practice, the characterisation of such a tester is out of reach and so we are faced with the same issue of instrument error. 

Engineering instruments allow us to perform full-process tensor tomography, provided we confine to a small number of time steps. This would indeed be very useful for estimating non-Markovianity and other correlations that are not readily available in restricted tomography. However, we can also put the engineered instruments to work for performing the temporal classical shadow tomography, which will allow us access a large number of multi-time correlations. Importantly, the latter contains the former as a limiting case. Thus we confine our focus to temporal classical shadow tomography.

\section{Correlated Noise Characterisation}
\label{sec:correlated-noise}

Currently, there exists a dearth of \textit{quantum characterisation verification and validation} {(QCVV)} tools for the characterisation of correlated noise. There are approaches at a coarse-grained level, with, for example, noise twirled into depolarising channels~\cite{Harper2020,flammiaACES}. Such methods average out the detailed information.
Specifically, any coherent or correctable noise or clues as to the physical origin of the correlations is washed away by the twirling practice. This information is crucial to inform better control of a device, the attention paid to different fabrication practices, and the design of bespoke quantum error correction codes. 

PTT from Sec.~\ref{methods:PTT} can perform this task in principle, but the resource requirements are too large to be tractable over more than a small number of steps. If, rather than considering generic multi-time correlations, we are interested in collections of small correlated marginals of the dynamics, then the problem becomes much more feasible. We approach this problem using the classical shadows developed thus far. This allows us to estimate arbitrary groups of process tensor marginals. The resulting estimates then contain all information about the structure of different correlations between those times. As discussed in the previous section, this is conditioned on tracing out the remainder of the observed steps and is hence only with respect to the class of non-unital dynamics. 

The method of classical shadows from Sec.~\ref{sec:shadow} permits the determination of $k-$point correlation functions. Translating this to processes, we examine two questions: (i) what is the average case probability of temporally correlated Pauli noise at different times? and (ii) given the application of some operation at an earlier time, how do the dynamics of the system change at a later point? This is measured in terms of both total non-Markovianity and temporal entanglement. The former gives a coarse view on temporal correlations but is an important benchmark for the feasibility of quantum error correction. The latter is a fuller diagnosis of the device. It quantifies the complexity and nature of the noise and may be used for a more sophisticated form of quantum error mitigation, such as probabilistic error cancellation. In particular, here we shall employ the strategies introduced in Sec.~\ref{sec:process_marg} for efficiently determining marginals of a process, and subsequently combine this with a matrix product operator ansatz.

We demonstrate this approach on an IBM Quantum device, obtaining these detailed diagnostics across many different time steps from only a relatively small number of shots.
In this work, we perform experiments on devices with mid-circuit measurement capabilities. Note that the choice to use an ancilla is because these measurements typically have a long duration (1--5 $\mu$s). Deferring to an ancilla allows the system to continue participating in dynamics -- otherwise, the window is lost if measuring the system directly. We employ the bootstrapped {IC} basis using the procedure outlined in Sec.~\ref{sec:IC-control}, summarised in Fig.~\ref{fig:instrument-creation}. For the local unitary operation, random Clifford operations are applied to the system qubit between system-ancilla interactions. The circuits are therefore of the form given in Fig.~\ref{fig:instrument-creation}c, and for each of the $N$ outcomes, a snapshot of the form of Eq.~\ref{eq:bootstrapped-shadow} constructed to form a classical shadow $\mathtt{S}(\Upsilon_{k:0}, N)$.

\subsection{Correlated errors}

On \emph{ibm\_perth}, we apply randomised instruments using our complete basis (Fig.~\ref{fig:instrument-creation}b) to characterise an idle process. Our goal is to estimate the likelihood of correlated Pauli errors and the size of memory across two single-step channels. Moreover, we want to do this in a setting where the surrounding qubits are undergoing nontrivial dynamics to induce a complex environment. For this, we apply random $SU(4)$ gates on neighbouring qubits. This mimics a realistic environment for a qubit sitting idle during a computation. We do this for a seven-step process and compute the correlated dynamical maps from $t_i\rightarrow t_{i+1}$ and $t_j\rightarrow t_{j+1}$, as they were defined in Eq.~\eqref{PT-marg}. Recall that these are given by
\begin{gather}\label{eq:correrr}
\hat{\mathcal{E}}_{j+1:j;i+1:i} :=
\text{Tr}_{\bar{i},\bar{j}}[\Upsilon_{k:0}],   
\end{gather}
where $\bar{i},\bar{j}$ indicates the trace over all process tensor legs except for $\{\mathfrak{i}_i, \mathfrak{o}_i,\mathfrak{i}_j,\mathfrak{o}_j\}$. In other words, the underlying physical process is $\Upsilon_{k:0}$, which we interrogate with randomised instruments and use the classical shadow results to reconstruct the marginal processes given in the last equation. 

\begin{figure}[t]
	\centering
	\includegraphics[width=\linewidth]{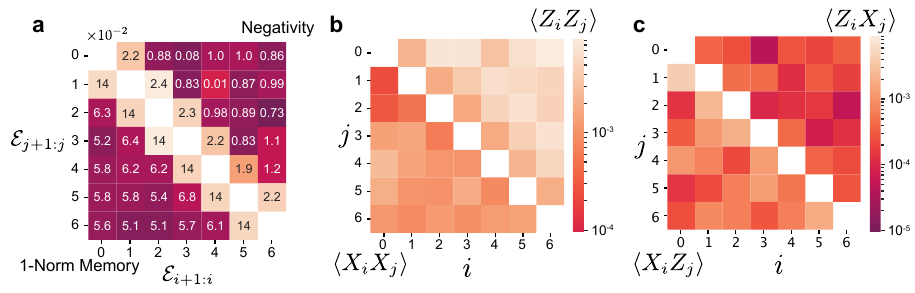}
	\caption[Heatmap capturing the two-map (four-body) correlated marginals on \emph{ibm\_perth} ]{The two-map (four-body) marginals of a system enduring naturally idle dynamics on \emph{ibm\_perth}. \textbf{a} We compute both the negativity between the two steps and the trace-distance between the maps and the product of their marginals. \textbf{b} We compute the joint overlap of correlated Pauli errors (their Choi states): $\langle \hat{P}_i\hat{P}_j\rangle - \langle \hat{P}_i\rangle\langle \hat{P}_j\rangle$ for $ZZ$ and $XX$ as well as \textbf{c} $ZX$ and $XZ$.}
	\label{fig:correlated-noise-maps}
\end{figure}

We collected $10^7$ shots of randomised instruments and then use Eq.~\eqref{eq:bootstrapped-shadow} to construct each of the corresponding classical shadows. The data is then further postprocessed to determine {IC} information about the marginals, and an {MLE} estimate is obtained. These estimates give us several representative components of the noise. Specifically, Figure~\ref{fig:correlated-noise-maps} displays:
\begin{itemize}
    \item The negativity between each pair of maps. That is, $\mathcal{N}_{j\mid i}[\hat{\mathcal{E}}_{j+1:j;i+1:i}]$, where $\mathcal{N}_{A\mid B}[\rho]:=\frac12 (\|\rho^{\Gamma_A}\|_1 - 1)$ is an entanglement monotone ($\Gamma_A$ is the partial transpose with respect to $A$), and can only be non-zero for genuinely quantum memory~\cite{Giarmatzi2021}.
    \item The trace distance $\frac12\|\hat{\mathcal{E}}_{j+1:j;i+1:i} - \hat{\mathcal{E}}_{j+1:j}\otimes \hat{\mathcal{E}}_{i+1:i}\|_1$, measuring distinguishability of the joint two steps from the tensor product of their marginals (which would be obtained, e.g., from quantum process tomography experiments).
    \item The connected Pauli probabilities across the two steps. That is, defining $\langle \hat{P}_i\hat{P}_j\rangle$ to be $\text{Tr}[\hat{\mathcal{E}}_{j+1:j;i+1:i}(\hat{P}_j\otimes\hat{P}_i)]$, where $\hat{P}$ is the Choi state of a Pauli operator, then we plot $\langle\hat{P}_i\hat{P}_j\rangle - \langle\hat{P}_i\rangle\langle\hat{P}_j\rangle$. 
\end{itemize}

Note that for marginals between non-neighbouring times, there is a maximal depolarising channel applied in the interim. However, we see more sensitivity here to the exact multi-time correlations, owing to the use of an {IC} probe to distinguish different aspects of the process. While we only considered a seven-steps process, the procedure scales only logarithmically with the number of times and so in principle could be extended far further.

Some of the results are consistent with device expectations, such as e.g. the prominence of correlated $Z$ errors compared to the other Paulis. Such errors are typically much more significant in solid-state devices compared to ones that change population. On the other hand, the size of the non-Markovianity shown in Figure~\ref{fig:correlated-noise-maps}a is surprisingly high. However, this does not translate into other correlated Pauli errors shown in Panels b and c. This is not surprising, as Panel a presents a basis-independent metric, whereas the latter Panels are basis-dependent. This suggests that correlated Pauli error models may not reveal the full picture of correlated noise, which our method can uncover.

\subsection{Dynamically Sampling Non-Markovian Open Quantum Systems}
\label{sec:dynamic-sampling}

We shall now move away from the context of noise characterisation and to the context of simulating non-Markovian processes. In recent years, there have been explosions of interest in using quantum devices to determine classically intractable properties. 
The seminal development of the randomised measurement toolbox provided not only an extremely practical {QCVV} tool, but kicked off a flurry of research poised at the question: what are the fundamental limits to which we can learn properties of quantum states? One recent development has been the idea to take quantum data from a device in the form of randomised measurements and process this with classical machine learning algorithms. It has been shown that this approach generates an exponential advantage in the learnability of quantum states. Meanwhile, Ref.~\cite{aloisio-complexity} showed the ability for quantum stochastic processes to be efficiently simulable (in the sampling sense) on a quantum computer, but intractable for a classical computer. This opens up the possibility of realising quantum advantage not only with large entangled registers but through small devices with complex environments.

In this section, we make a step towards this goal, by specifically considering the characterisation of multi-time sampling statistics in practice. This is hence a \emph{dynamic} sampling problem, whose complexity is studied in Ref~\cite{aloisio-complexity}. We move on to the study of much larger processes, coarser observations, and the evolution of marginals of a process. These fit under the broad umbrella of macroscopic process properties and can tell us, for example, how microscopic properties change as a function of time steps. Given the concerted effort to simulate correlated quantum processes classically~\cite{strathearn_efficient_2018, Luchnikov2019L, Cygorek-2022}, this connection lays the groundwork for fully general quantum simulation of non-Markovian open quantum systems. As well as simulating the dynamics (system as a function of time), a quantum computer may also access the dynamical multi-time correlations, which encapsulate every extractable piece of information from the system. 

We perform a demonstration that consolidates our tools and reconstructs a long non-Markovian process simulated on a quantum device. Specifically, we consider one end of a Heisenberg-coupled spin-chain and estimate the 20-step/42-body process tensor Choi state using the \textit{finitely correlated state} ansatz~\cite{PhysRevLett.111.020401}. This method reconstructs the global quantum state, within the finitely correlated approximation (i.e., that the correlation structure is one dimensional and exponentially decreasing with distance), using only the marginals of an odd number of neighbouring spins that satisfy certain left and right invertibility conditions. More specifically, this compact representation is a subclass of matrix product operators (MPOs). Further information on MPOs and other tensor network states may be found in Ref.~\cite{Cirac2020MatrixPS}.
As an ansatz, to describe an $n$-body state, choose left and right ranks to be $d^{2l}$ and $d^{2r}$ respectively (where left and right ranks are the dimensions of the image of a matrix acting as a map to the left or right, respectively). Here, $d$ is the dimension of each local body. This requires $R$-body marginals with $R = l + r + 1$, leading to a total of $n - R + 1$ local marginals. In the process tensor context, we start with $ k$-step process, represented by a $2k+1$-body state. If $l = r :=\ell$, then the problem requires estimation of all $\ell$ step marginals: $\{\Upsilon_{\ell:0}$, $\Upsilon_{\ell+1:1}$, $\cdots$, $\Upsilon_{k:k-\ell}\}$. With the input of classical shadows, these $k - \ell + 1$ marginals may be simultaneously estimated, leaving the characterisation scaling logarithmically in the number of steps. 

\begin{figure}[t]
	\centering
	\includegraphics[width=0.6\linewidth]{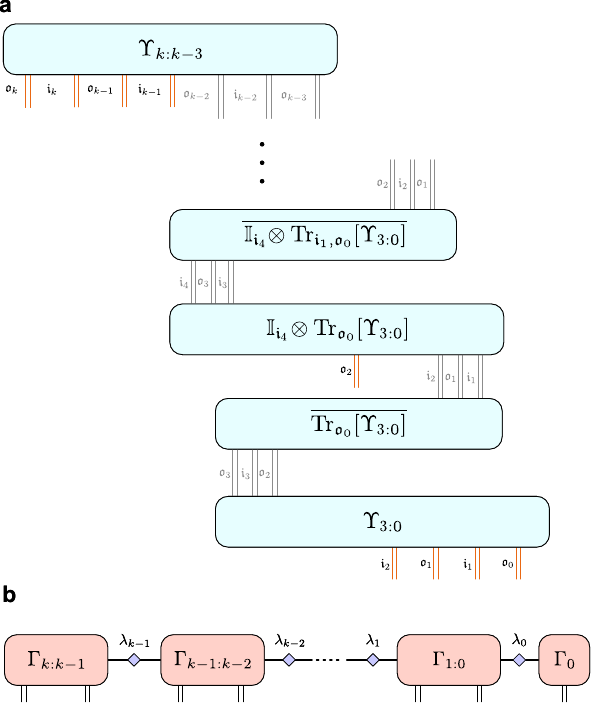}
	\caption[Graphical illustration of a reconstructed many-body process tensor from its marginals ]{Graphical illustration of a reconstructed many-body process tensor from its marginals. \textbf{a} We construct the process tensor as a finitely correlated state by stitching together $\ell$ step marginals estimated from classical shadows. Here is an example taking $\ell=3$. Note that the overline denotes a pseudoinverse of the matrix. Orange legs are free indices, and grey legs are contracted indices between process tensor marginals. \textbf{b} The resulting many-time representation after condensing down the above, and appropriately splitting sites at the start and finish. The $\lambda_j$ indicate summation over the indices given in the marginal pseudoinverses. }
	\label{fig:PT-MPO}
\end{figure}

Full details of the theorem reconstructing a state from its marginals may be found in Ref.~\cite{PhysRevLett.111.020401}, however, we reconfigure the statements here in the form of a graphical diagram in Figure~\ref{fig:PT-MPO} to reconstruct a \textit{finitely correlated process ansatz}. The procedure involves computing pseudoinverses to connect together process tensor marginals. This further motivates the need for a robust marginal estimate, since inverses can only be computed with complete tomographic information about a state. A description of the Ref.~\cite{PhysRevLett.111.020401} procedure as adapted to process tensors is as follows. We start with $\Upsilon_{\ell:0}$ and move one leg across to obtain the next $2\ell+1$ body marginal $\text{Tr}_{\mathfrak{o}_{\ell+1}, \mathfrak{o}_0}[\Upsilon_{\ell+1:0}]$. However, causality constraints mean that this term reduces to $\mathbb{I}_{\mathfrak{i}_{\ell+1}} \otimes \text{Tr}_{\mathfrak{o}_0}[\Upsilon_{\ell:0}]$, which is information we already have. Moving one leg over, the next marginal is $\Upsilon_{\ell+1:1}$. This continues until we get to $\Upsilon_{k:k-\ell}$. Collectively, these form the set
\begin{equation}
		E:=\{\Upsilon_{k:k-\ell}, \mathbb{I}_{\mathfrak{i}_k}\otimes \text{Tr}_{\mathfrak{o}_{k-\ell-1}}[\Upsilon_{k-1:k-\ell-1}], \Upsilon_{k-1:k-\ell-1},\cdots,\mathbb{I}_{\mathfrak{i}_{\ell+1}}\otimes\text{Tr}_{\mathfrak{o}_0}[\Upsilon_{\ell+1:0}],\Upsilon_{\ell:0}\}
\end{equation}
We will contract this set with the set of their pseudoinverses, which we denote with an overline as $\overline{\Upsilon_{\ell:0}}$.

The connections of these terms to construct the overall finitely correlated process is depicted in Figure~\ref{fig:PT-MPO}a, with the effective many-time state depicted in Figure~\ref{fig:PT-MPO}b. The net result is an MPO of the form
\begin{gather}\label{eq:fcs}
    \Upsilon^{(fcp)}_{k:0} = \sum_{\lambda_0,\ldots,\lambda_{k-1}} \Gamma_0^{(\lambda_0)}  \Gamma_{1:0}^{(\lambda_0,\lambda_1)}  \cdots 
    \Gamma_{k-1:k-2}^{(\lambda_{k-2},\lambda_{k-1})}   \Gamma_{k:k-1}^{(\lambda_{k-1})}.
\end{gather}
The $\Gamma$ matrices are the MPO tensors representing quantum channels between neighbouring steps. For example, $\Gamma_{j+1:j}$ has the physical indices $\mathfrak{i}_j,\mathfrak{o}_j$. The $\lambda$s represent the bond of the MPO, with the maximum size of any $\lambda$ being the bond dimension. More specifically, the MPO bond dimension corresponds to the size of the effective requirement which would be required to carry forward temporal correlations, and is hence a measure of non-Markovianity~\cite{Pollock2018}.

\subsection{Reconstructing finitely correlated process ansatz on NISQ devices}

\begin{figure}
	\centering
	\includegraphics[width=0.5\linewidth]{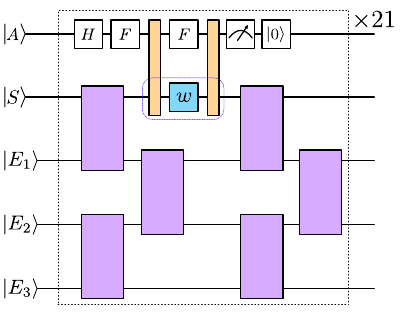}
	\caption[Circuit diagram collecting multi-time statistics as part of a spin chain ]{Circuit diagram collecting multi-time statistics as part of a spin chain. An exchange interaction is driven between neighbouring qubits across 21 repeated blocks, and ancilla qubit used to probe the system. A $V_{\pi/4}$ operation is used to interact $S$ and $A$, followed by projective measurement on $A$. In each run the local unitary on $S$ is chosen to be a random Clifford operation, generating the classical shadow data. }
	\label{fig:spin-chain-circ}
\end{figure}

We use finite correlated process ansatz as a platform to experimentally reconstruct correlated multi-time statistics for large processes. Specifically, we use a quantum device to simulate a spin chain and then examine the properties of the quantum stochastic process at one end of the chain. The device used was \emph{ibm\_lagos}, where the spin chain was simulated with four qubits, see Figure~\ref{fig:spin-chain-circ}. One qubit is treated as the systems $S$, and the rest as environment. An additional ancilla qubit was also used for the purpose of applying quantum instruments to the system. At each step, an exchange coupling is simulated by the application of a random $XX + YY + ZZ$ rotation between each of the neighbouring qubits. Using the ancilla qubit, a random (bootstrapped, as per Sec.~\ref{sec:IC-control}) instrument is applied to the system for a total of $4.4\times 10^7$ shots. Eq.~\eqref{eq:bootstrapped-shadow} is then used to construct the classical shadow for each shot, and the data post-processed to produce the three step marginals $\{\Upsilon_{l:l-3}\}_{l=3}^{21}$. Then, by evaluating multi-time expectation values we can generate a predicted probability distribution for time-series data -- the 21 measurements across the circuit. The overall circuit diagram is given in Figure~\ref{fig:spin-chain-circ}.

To validate that the estimate does indeed accurately capture environmental dynamics, 170 different sequences of random instruments are then run on the device at $16\:384$ shots. The outcomes of the instruments across the 21 different times are collected and used to generate different marginal probability distributions. The effect of marginalising over the outcomes at a given time is that the operation applied is a deterministic {CPTP} channel. The marginal probability distributions for each sequence of operations is predicted by our process characterisation and compared via Hellinger fidelity to the actual results. 

\begin{figure}
	\centering
	\includegraphics[width=\linewidth]{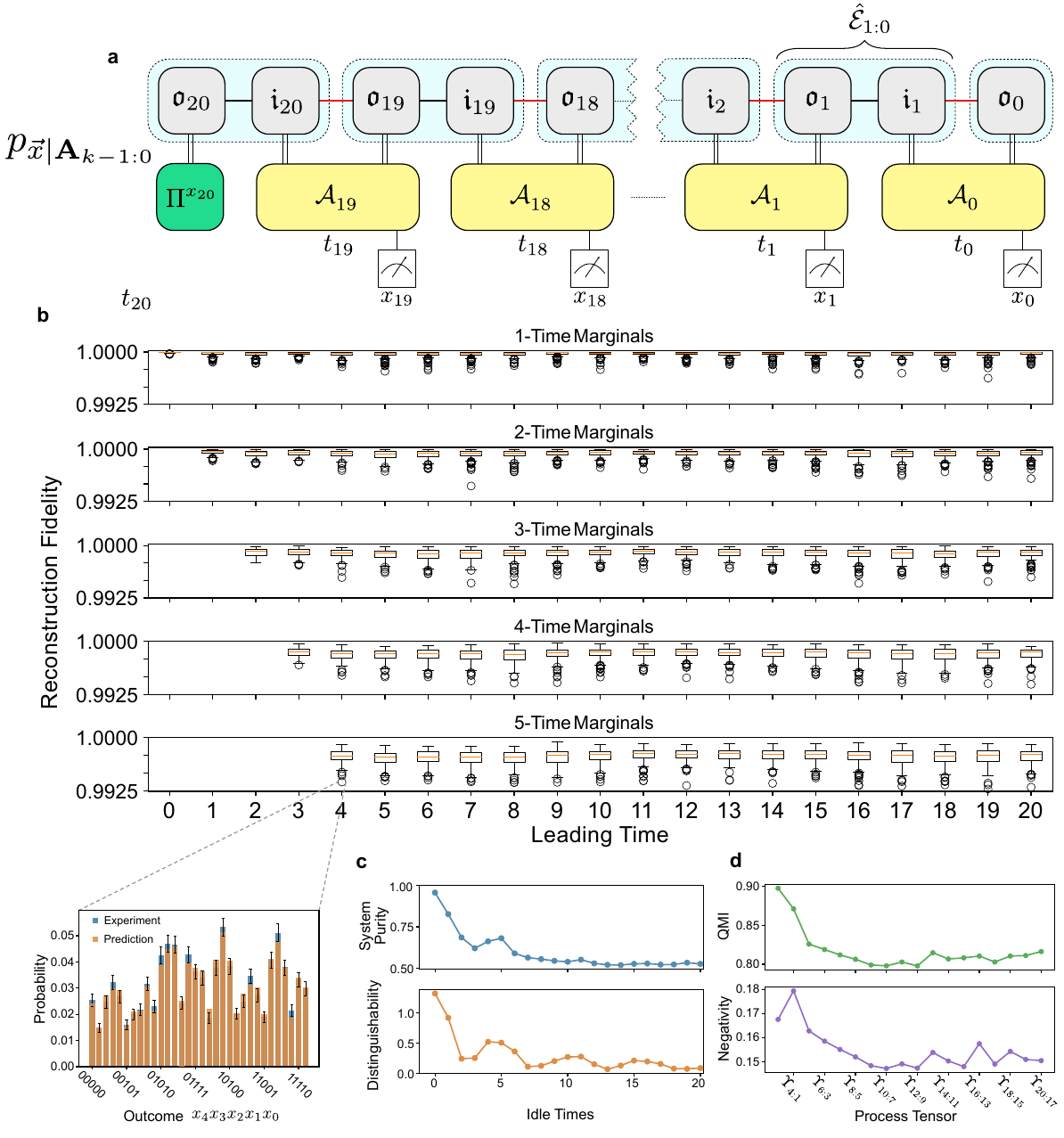}
	\caption[Experimentally reconstructing dynamic sampling statistics from a finitely correlated process tensor ]{
	\textbf{a} We characterise a 20-step process of a non-Markovian quantum environment on \emph{ibm\_lagos} and construct its finitely correlated state representation, in Eq.~\eqref{eq:fcs}. \textbf{b} This characterisation is then validated against 170 sequences of random instruments: the reconstructed process tensor is used to predict different $l$-time marginal probability distributions of the instrument outcomes and compared against the experimental results.
	\textbf{c} Properties of the system when left idle as a function of time step, showcasing conventional measures of non-Markovianity. We observe the non-monotonic behaviour of both the purity of the system, and the trace distance between two states initialised as $Z$ and $Y$ eigenstates. \textbf{d} Properties of the 3-step marginal process tensors as a function of time, we measure both the total memory and the temporal quantum entanglement, demonstrating a genuinely complex non-Markovian process.}
	\label{fig:full-mpo-results}
\end{figure}

The cross-validation of multi-time statistics is as follows. Consider one such test sequence of random instruments $\mathbf{J}_{20:0} = \{\mathcal{J}_{20}^{(x_{20})},\mathcal{J}_{19}^{(x_{19})},\cdots \mathcal{J}_0^{(x_0)}\}$. Note that the last instrument, $\mathcal{J}_{20}$, is actually a {POVM} -- an instrument where the post-measurement state is thrown away. Each instrument is selected by choosing a Haar random unitary to act on the system, and the ancilla qubit measurement outcomes recorded. Because of the characterisation procedure given in Sec.~\ref{methods:PTT}, each instrument's effective map is known, regardless of the local unitary. A single circuit consequently induces a joint probability distribution
\begin{equation}
	\label{eq:prob-dist}
	\mathbb{P}(x_{20},t_{20}; x_{19},t_{19};\cdots;x_0,t_0 | \mathbf{J}_{20:0})
\end{equation}
for the measurement outcomes at the different times. We first sample from this distribution for a given $\mathbf{J}_{20:0}$ by running the sequence on the device at $16\:384$ shots and recording each of the results. Naturally, at $2\times 10^6$ outcomes, this distribution is already too large to sensibly measure and compare. Hence, we consider marginals of the distribution. A marginal of length $\ell$ at a leading time $j$ is the marginal distribution of Eq.~\eqref{eq:prob-dist} 
\begin{equation}
	\label{eq:prob-marg}
	\mathbb{P}(x_{j},t_{j}; x_{j-1},t_{j-1};\cdots;x_{j-\ell+1},t_{j-\ell+1} | \mathbf{J}_{j:0}).
\end{equation}

The estimate of the finitely correlated state representation of $\Upsilon_{20:0}$ can produce a corresponding estimate for the distributions in Eq.~\eqref{eq:prob-marg} by computing
\begin{align}
&	\label{eq:PT-probs}
	p_{x_jx_{j-1}\cdots x_{j-\ell+1} | \mathbf{J}_{j:0}} = \\ \nonumber &\quad
\text{Tr}\left[\left(\mathbb{I}_{t_{20}:t_{j+1}}\otimes \mathcal{J}_j^{(x_j)\text{T}}\otimes \mathcal{J}_{j-1}^{(x_{j-1})\text{T}}\otimes\cdots \otimes \mathcal{J}_{j-\ell+1}^{(x_{j-\ell+1})\text{T}}\otimes \mathcal{J}_{j-\ell-2}^{\text{T}}\otimes \cdots \otimes \mathcal{J}_0^{\text{T}}\right)\cdot \Upsilon_{20:0}\right],
\end{align}
which, for the resulting graphical diagram in Figure~\ref{fig:PT-MPO}, can be contracted efficiently. Note that where we have dropped the instrument superscript, this indicates that we have discarded the ancilla outcome, effecting a {CPTP} map: $\mathcal{J}_i = \mathcal{J}_i^{(0)} + \mathcal{J}_i^{(1)}$. The boxplots of Figure~\ref{fig:full-mpo-results}b display the Hellinger distance between the quantum computer sample, and Eq.~\eqref{eq:PT-probs} for each of the 170 sequences, for each marginal $1-5$, and for each time $0-20$.

In the last two panels of we examine the non-Markovian properties of the process from different perspectives. Figure~\ref{fig:full-mpo-results}c measures the purity $\Tr[\rho^2]$ of the idle state of the system as a function of time -- with no control operations applied. The oscillations seen indicate decoherence and recoherence of the state, a strong characteristic of information backflow from the environment~\cite{rivas-NM-review}. 
We also demonstrate another aspect of two-time correlations, which is the non-monotonicity of trace distance between two differently initialised states~\cite{PhysRevA.83.052128}. We start $\rho_1 = |0\rangle\!\langle 0|$ and $\rho_2 = |i+\rangle\!\langle i+|$, and plot $\|\rho_1(t) - \rho_2(t)\|_1$ for each time step. This quantity oscillates in phase with the state purities, and exhibits a much stronger signal. Collectively, these demonstrate substantial environmental back-action. Thus, the IBM Quantum device is able to coherently simulate this non-Markovian dynamics, which we have demonstrated we have been able to fully characterise.

 Meanwhile Figure~\ref{fig:full-mpo-results}d conveys exact measures of non-Markovianity throughout the process. We see that both the generalised {QMI}, and negativity across the middle cut of the three-step process marginals both decay to a non-zero constant value. Despite the nominal decay of the system, memory effects persist in the dynamics, as can be detected through mid-circuit measurements. The reason the memory persists in the multi-time statistics but not in the purity or distinguishability measures is two-fold: (i) these measures are not as general, and will always lower bound the multi-time quantity, and (ii) measuring the state of the system as a function of time is likely to lead to dissipation. Incorporating the effects of control operations can act as a randomiser of the state of the system, and effectively decoupling from the decoherence process.
 
 Here, we have gone beyond the state-of-the-art in characterising quantum stochastic processes. We have shown how one may completely characterise arbitrarily many time-steps in a quantum process by using the available hardware to manufacture full system control. Through our demonstration we validate the ability to predict the statistics produced by arbitrary sequences of instruments with mid-circuit measurements. This demonstrates the first full-scale characterisation of a quantum stochastic process, and showcases how multi-time statistics can be learned in practice.
The memory of the process studied is genuinely complex, but not so large that our finitely correlated state model is unable to capture the necessary degrees of freedom. This further provides a clue on constructing compressed models to effectively determine non-Markovian dynamics. Our approach and demonstration broadens the study of processes to a far more macro level, widening the scope to more intricate considerations such as thermalisation and equilibration.

Because we have simulated a relatively small environment, and because noise on quantum devices is expected to render this mostly dissipative, we take three step marginals as our ansatz for reconstruction. This assumes that the left and right ranks at each site of the process tensor is no more than 64. With the three-step marginals estimated through repeated randomised measurements made on the ancilla qubit, we may now reconstruct the entire 20-step process tensor Choi state according to our finite correlation ansatz. It should be stated that that this reconstruction falls into the realm of classically simulable, the intended purpose is to show that dynamical sampling of multi-time statistics is possible even with near-term devices. Moreover, one use the procedure and apply it to an instance where the environment is nontrivial. This not only establishes that scalable non-Markovian characterisation is achievable but presents a new aspect to the simulation capabilities of quantum computers.

\section{Discussion}
\label{sec:conc}
Although the study of many-body physics has uncovered a rich tapestry of structure in different corners of applied quantum mechanics, temporal quantum correlations remain relatively under-explored. In this work, we have motivated, demonstrated, and made accessible the study of many-time physics beyond the two-time correlators regime. Multi-time correlations will showcase the full potential of the field of many-time physics, including capturing dynamical phases of matter~\cite{heyl2018dynamical}.

To capture multi-time correlations we first generalised the theory of classical shadow tomography to the temporal scenario. However, implementing temporal shadow tomography requires nontrivial instruments. We then make use of ancilla qubits, to realise these instruments on real quantum hardware. Combining these two innovations enabled us to access a variety of non-Markovian phenomena, as well as other multi-time observables. This allowed us to show correlated Pauli errors do not fully capture the size on non-Markovian correlations. Next, we employed the finitely correlated state ansatz to reconstruct a large scale quantum process on a quantum processor.

Rather than merely simulating non-Markovian quantum processes on fully controllable quantum computers, this allows observation of complex naturally occurring phenomena~\cite{PhysRevA.61.023603, jaksch2005cold,PhysRevLett.101.260404,PhysRevA.90.032106,caruso2009highly,lambert2013quantum}. Our results on IBM Quantum devices show that even idle qubits can exhibit surprisingly complex and temporally correlated dynamics. We have discovered that the non-Markovianity persists not simply as a classical set of correlations, but has genuine quantum properties such as temporal quantum entanglement. This prompts the need for further study of non-Markovian behaviour on {NISQ} devices. Specifically, if not fabricated away, correlated noise ought to be able to be converted into clean channels through appropriate control sequences~\cite{berk2021extracting}.

We have also shown how to arbitrarily capture many-time physics through repeated probes of a system. This naturally fits into models of quantum simulation, but provides strictly more information than simply observing evolution of a state. From this, we can envisage the transduction of multi-time observables from a single non-Markovian quantum system into (classical or quantum) machine learning algorithms, which has recently been shown as a promising avenue for quantum advantage~\cite{Huang2021}. 

Our methods will find utility in both the search for fundamental physics and building better quantum hardware. The former comes as quantum simulation of open quantum dynamics, which not fully understood at the multi-time scale. For the latter, our tools can help with better quantum error characterisation and mitigation, and building a pathway for quantum error correction in presence of correlated noise.

\begin{acknowledgments}
We thank Felix Pollock for valuable discussions. This work was supported by the University of Melbourne through the establishment of an IBM Quantum Network Hub at the University.
G.A.L.W. was supported by an Australian Government Research Training Program Scholarship during the time of this research. 
K.M. is supported through Australian Research Council Future Fellowship FT160100073.
K.M. and C.D.H. acknowledge the support of Australian Research Council's Discovery Project DP210100597.
K.M. and C.D.H. were recipients of the International Quantum U Tech Accelerator award by the US Air Force Research Laboratory.
\end{acknowledgments}

\providecommand{\newblock}{}

\end{document}